\documentclass[superscriptaddress,amsmath,amssymb,aps,twocolumn]{revtex4-1}

\usepackage{newtxtext,newtxmath}
\usepackage[T1]{fontenc}
\usepackage[pdftex]{graphicx}
\usepackage{bm}
\usepackage{textcomp}
\usepackage{braket}
\usepackage{siunitx}
\usepackage{enumitem}
\usepackage{caption}

\usepackage{times}
\usepackage{dcolumn}
\usepackage{amsmath}
\usepackage[dvipsnames]{xcolor}
\usepackage[labelsep=space]{caption}
\usepackage{setspace}
\usepackage{amssymb}
\usepackage[version=3]{mhchem}

\begin{document}

\title{A review of dielectric optical metasurfaces for wavefront control}

\author{Seyedeh Mahsa Kamali}
\thanks{These two authors contributed equally to this work.}
\affiliation{T. J. Watson Laboratory of Applied Physics and Kavli Nanoscience Institute, California Institute of Technology, 1200 E California Blvd., Pasadena, CA 91125, USA}
\author{Ehsan Arbabi}
\thanks{These two authors contributed equally to this work.}
\affiliation{T. J. Watson Laboratory of Applied Physics and Kavli Nanoscience Institute, California Institute of Technology, 1200 E California Blvd., Pasadena, CA 91125, USA}
\author{Amir Arbabi}
\affiliation{Department of Electrical and Computer Engineering, University of Massachusetts Amherst, 151 Holdsworth Way, Amherst, MA 01003, USA}
\author{Andrei Faraon}
\email{Corresponding author: A.F: faraon@caltech.edu}
\affiliation{T. J. Watson Laboratory of Applied Physics and Kavli Nanoscience Institute, California Institute of Technology, 1200 E California Blvd., Pasadena, CA 91125, USA}

\begin{abstract}
During the past few years, metasurfaces have been used to demonstrate optical elements and systems with capabilities that surpass those of conventional diffractive optics. Here we review some of these recent developments with a focus on dielectric structures for shaping optical wavefronts. We discuss the mechanisms for achieving steep phase gradients with high efficiency, simultaneous polarization and phase control, controlling the chromatic dispersion, and controlling the angular response. Then we review applications in imaging, conformal optics, tunable devices, and optical systems. We conclude with an outlook on future potentials and challenges that need to be overcome.
\end{abstract}

\maketitle

\section{Introduction}
Optical metasurfaces are two-dimensional arrays of subwavelength scatterers that are designed to modify different characteristics of light such as its wavefront, polarization distribution, intensity distribution, or spectrum~\cite{Holloway2012IEEEAntPropMag,Kildishev2013Science,Yu2014NatMater,Estakhri2016JOSAB,
Jahani2016NatNano,Lalanne2017LaserPhotonRev,Staude2017NatPhoton,Genevet2017Optica,
Hsiao2017SmallMeth,Qiao2018AdvOptPhoton,ZiJing2017JOpt,Kruk2017ACSPhoton}. The subwavelength scatterers (referred to as meta-atoms in this context), capture and reradiate (or scatter) the incident light. Depending on the meta-atom design, the scattered light might have different characteristics compared to the incident light. For instance, it might have a different phase, polarization ellipse, angular distribution, intensity, and/or spectral content. For most metasurfaces, the output is either the scattered light or the interference between the scattered and the incident light. By proper selection of the meta-atoms and their locations in the array, the characteristics of light interacting with the metasurface can be engineered. As a result, different conventional optical components such as gratings, lenses, mirrors, holograms, waveplates, polarizers and spectral filters may be realized. Furthermore, a single metasurface might provide a functionality that may only be achieved by a combination of conventional optical components~\cite{Arbabi2015NatNano} or an entirely novel functionality~\cite{Kamali2017PRX}. 
Typically metasurface optical components are subwavelength thick, have a planar form factor, and can be batch-fabricated at potentially low cost using the standard micro and nano-fabrication processes. In the past few years, the efficiency of the optical metasurfaces has improved significantly by switching from metallic (or plasmonic) meta-atoms to high refractive index dielectric ones. The combination of the relatively high efficiency, potentially low cost, and the planar and thin form factor has generated significant interests in metasurface optical components, and has attracted a large number of researchers from different disciplines and with different backgrounds. The result has been the rapid expansion of the field. Here we discuss the recent progress in the development of optical metasurfaces, focusing on dielectric metasurfaces that modify the wavefront and/or polarization distribution of light. We primarily discuss work done by the authors, and its relation to other work in the field.

Optical metasurfaces are conceptually similar and are technically closely related to the reflectarrays and transmitarrays which have been studied for decades in the microwave community~\cite{Huang2007Reflectarray}. For example, the idea of using elements (or meta-atoms) with different sizes and shapes has been used as early as 1993 in that community for creating spatially varying phase profiles~\cite{Pozar1993ElectronLett}. Early demonstrations of optical metasurfaces which used metallic meta-atoms are similar to their microwave counterparts~\cite{Yu2011Science}. Another example is the use of the geometric (or Pancharatnam-Berry) phase for controlling the wavefront of circularly polarized waves that also has been used before in the microwave community~\cite{Malagisi1978EASCON,Huang1998IEEETAP}. In addition, many of the properties, design techniques, and models for metasurface components are similar to those used in the context of diffractive optical elements (DOEs). The recognition of these similarities can be beneficial in the design and development of metasurfaces. An example of the results that are similarly applicable to metasurfaces and DOEs, is the ray optics treatment of the refraction, reflection, and transmission of rays upon interaction with surfaces that impart a spatially varying phase. This topic has been well studied in diffractive optics and the resulting relation is known as the grating equation. A general treatment of this problem when the phase imparting surface has an arbitrary curved shape can be found in~\cite{Fairchild1982OptEng}. Depending on how they are realized, DOEs have different categories including kinoforms~\cite{Lesem1969IBMJRD}, holographic optical elements, computer-generated holograms~\cite{O'Shea2004}, and effective medium structures~\cite{Stork1991OptLett,Farn1992ApplOpt}, and are realized using structures that are different from metasurfaces. However, similar to metasurfaces, they impart spatially varying phase excursion and are modeled as spatially varying phase masks. As a result, many of the techniques, theories, and designs developed for and using DOEs are directly applicable and transformable to metasurfaces. Some of the examples include the algorithms for the design of phase profiles that project desired intensity patterns~\cite{O'Shea2004},  estimations of diffraction efficiency for quantized phase levels~\cite{Swanson1991}, elimination of the spherical aberrations by proper selection of the phase profile~\cite{Swanson1989}, elimination of coma aberration of a phase profile by applying it on a curved spherical surface~\cite{Welford1973OptCommun}, and removal of other monochromatic aberrations by using a stop~\cite{Buralli1991AppOpt} or multiple cascaded elements~\cite{Buralli1989AppOpt}.

One of the advantages of recognizing the relation of the metasurfaces to other DOEs, is the identification of the potential advantages of the metasurfaces over conventional DOEs. For instance, low-cost, efficient, and relatively wideband diffractive lenses can be realized using conventional DOEs~\cite{rpcphotonics} but their performance degrades significantly with increasing their numerical apertures (NAs). As we discuss in section~\ref{High-efficiency high-gradient phase control}, properly designed metasurfaces can outperform conventional DOEs. Chromatic aberration is another example of challenges that both DOEs and metasurfaces face. Similar to DOEs, the chromatic aberration of metasurfaces is caused by the phase discontinuities at the zone boundaries \cite{Arbabi2016Optica}. As we discuss in section~\ref{Controlling chromatic dispersion}, metasurfaces provide multiple solutions for mitigating the chromatic aberration.

\section{Recent developments}

\begin{figure*}[htbp]
\centering
\includegraphics[width=1\linewidth]{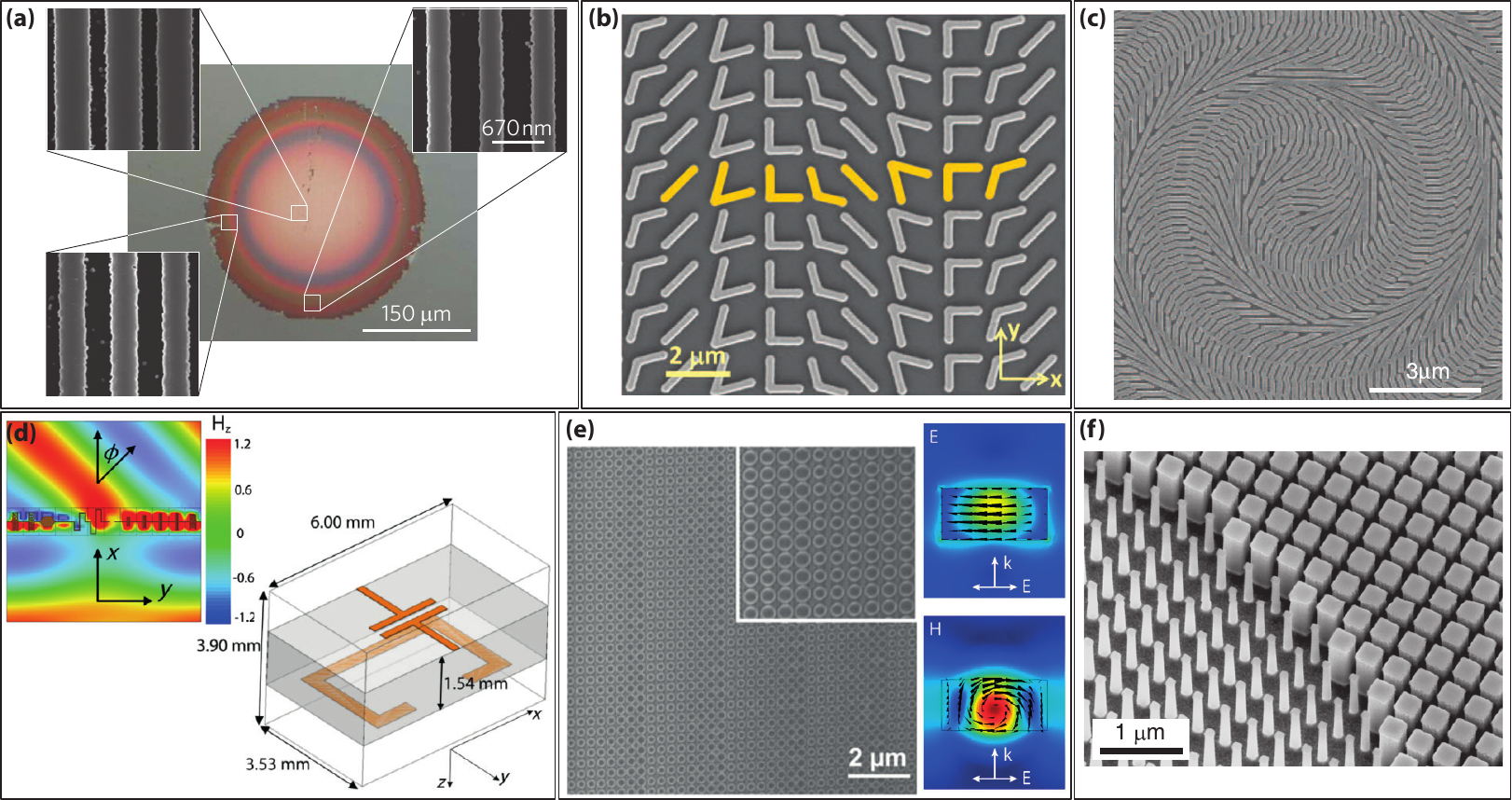}
\caption{\textbf{Recent advances in metasurfaces}. \textbf{(a)} Optical and scanning electron micrographs of a high-contrast grating mirror with focusing ability~\cite{Fattal2010NatPhoton}. \textbf{(b)} Scanning electron micrograph of a plasmonic metasurface beam deflector~\cite{Yu2011Science}. \textbf{(c)} A geometric phase axicon with dielectric microbars~\cite{Lin2014Science}. \textbf{(d)} Microwave beam deflection with a metallic Huygens' metasurface along with unit cell of the beam deflector~\cite{Pfeiffer2013PhysRevLet}. \textbf{(e)} Scanning electron micrograph of a dielectric Huygens' beam deflector. Simulated field intensities are plotted on the right~\cite{Yu2015LaserPhotonRev}. \textbf{(f)} Scanning electron micrograph of a portion of a high-contrast transmitarray lens~\cite{Kamali2016LaserPhotonRev}.}
\label{Fig1_Recent}
\end{figure*}

In the past few years, the advances and the wider accessibility of micro and nano-fabrication technologies, along with an increased interest in dielectric high-contrast~\cite{Fattal2010NatPhoton,Lu2010OptExp,Fattal2011AdvPhoton} and plasmonic structures~\cite{Yu2011Science,Zhao2011PhysRevB,Zhao2011Metamat,Aieta2012NanoLettAberr,Huang2012NanoLett} for manipulation of optical wavefronts, have caused a surge in the research field of metasurfaces. Two of the early works using high contrast mirrors and plasmonic scatterers are shown in Figs.~\ref{Fig1_Recent}a and \ref{Fig1_Recent}b, respectively. The ultrathin form factor of plasmonic structures, and the great interest in the field of plasmonics itself, resulted in most of the earlier works using a single metallic layer to manipulate light using resonance phase, geometric phase, or their combination~\cite{DiFalco2011APL,Aieta2012NanoLettAberr,Yu2012NanoLett,Kildishev2013Science,Li2013NanoLett,
Ni2013NatCommun,Pors2013OptExp,Pors2013NanoLett}. However, material losses and fundamental limitations of single layer thin plasmonic metasurfaces (especially in the transmission mode)~\cite{Datthanasombat2001APSIS,Monticone2013PhysRevLett,Arbabi2017SciRep} significantly limit their performance. Dielectric geometric phase elements based on nano-beam half waveplates (similar to the example shown in Fig.~\ref{Fig1_Recent}c) have also been investigated~\cite{Hasman2003ApplPhysLett,Lin2014Science} for wavefront shaping. These elements are designed to work with one polarization, and achieving simultaneously both high efficiency and large deflection angles is challenging because of significant coupling between the elements.

To overcome the fundamental limitations of ultra-thin metasurfaces, Huygens' metasurfaces were introduced~\cite{Pfeiffer2013PhysRevLet} that allow for simultaneous excitation of modes with equal electric and magnetic dipole moments. These structures do not have a deep subwavelength thicknesses, as shown in Fig.~\ref{Fig1_Recent}d where the wavelength is 30~mm. Despite their success in lower frequencies~\cite{Epstein2014PhysRevB,Epstein2014IEEETAP,Kim2014PRX,Jia2015AIPAdv,
Epstein2016NatCommun,Epstein2016JOSAB}, in the optical domain metallic Huygens' metasurfaces are still limited by material losses and often require complicated fabrication. As a result, dielectric Huygens' metasurfaces were explored~\cite{Staude2013ACSNano,Yu2015LaserPhotonRev,Campione2015OptExp,
Decker2015AdvOptMat,Asadchy2016JOSAB,Zhao2016SciRep,Forouzmand2017AdvOptMat} that allowed for two longitudinal resonance modes with dominant electric and magnetic dipole moments with the same frequency to circumvent material losses [Fig.~\ref{Fig1_Recent}e]. There are, however, some challenges that limit the practicality of dielectric Huygens' metasurfaces. First, full 2$\pi$ phase coverage at a single wavelength, which is what matters for wavefront manipulation, while keeping a high transmission requires changing all sizes of the resonators (including their heights) which is challenging to achieve with the conventional planar microfabrication technology. Second, the coupling between adjacent meta-atoms is considerable in Huygens' metasurfaces and this significantly degrades the performance of devices with large deflection angles as they require fast varying structures~\cite{Yu2015LaserPhotonRev}. As a result, more groups started investigating the High-Contrast transmit/reflect Array (HCTA/HCRA or HCA) structures (similar to the one shown in Fig.~\ref{Fig1_Recent}f) that use thicker (about 0.5$\lambda$ to $\lambda$) high-index layers to pattern the metasurface~\cite{Fattal2011AdvPhoton,ArbabiCLEO2014,Vo2014IEEEPhotonTechLett,
Arbabi2015NatCommun,Arbabi2015NatNano,Arbabi2015OptExp,Zhan2016ACSPhotonics,Kruk2016APLPhotonics,
Khorasaninejad2016Science,Paniagua2018NanoLett,Chen2017NanoLett,Wang2016NanoLett}. These structures are very similar to the blazed binary optical elements that are at least two decades old~\cite{Astilean1998OptLett,Lalanne1998OptLett,Lalanne1999JOSAA_Multimode,Lalanne1999JOSAA}, nevertheless, they outperform other classes of metasurfaces in many wavefront manipulation applications. In the following, we will review the areas where metasurfaces have demonstrated wavefront control capabilities beyond those of conventional diffractive optical elements, and then discuss a few application areas where metasurfaces can be employed.

We should note here that the applications of optical metasurfaces in the general sense of the word (i.e., patterned thin layers on a substrate) go beyond spatial wavefront manipulation. Thin light absorbers~\cite{Mosallaei2005IEEEAntPropSymp,Martinez2013EuCAP,Yao2014NanoLett,Jung2015IEEEPJ,
Yang2015APMC,Radi2015ACSPhoton,Azad2016SciRep,Guo2016OptExp,Kim2016NanoLett,Luo2016ApplPhysLett,
Wan2016NanoEnergy,Jung2017SciRep,Li2017IEEEMTT,Sun2017AIPAdv,Tang2017IEEEPTL}, optical filters~\cite{Magnusson1993Patent,Shin1998OptEng,Shokooh2007OptLett,Ortiz2011APSURSI,
Ellenbogen2012NanoLett,Ortiz2013IEEEMWCL,Wang2015IEEEJSTQE,Horie2015OptExp,
Horie2016OptExp,Horie2017NanoLett,Yue2016SciRep,Yamada2017OptLett,Limonov2017NatPhoton},  nonlinear~\cite{Lee2014Nature,Tymchenko2015PhysRevLett,Wolf2015NatCommun,Camacho2016NanoLett,Nookala2016Optica,
JafarZanjani2016AIPAdv,Liu2016NanoLett,Makarov2017NanoLett}, and anapole metasurfaces~\cite{Miroshnichenko2015NatCommun,Wu2018ACSNano} are a few examples of such elements. This review is focused on applications of metasurfaces in wavefront manipulation, and therefore it doesn't cover these other types of metasurfaces.

\section{New capabilities for controlling light enabled by metasurfaces}
In this section, we review the recent findings and achievements using metasurfaces that include devices providing functionalities not achievable with conventional diffractive optical elements. We will discuss their ability to bend light by large angles with high efficiency, completely control phase and polarization of light, engineer the chromatic dispersion of optical elements, and independently control the function of a device for different illumination angles.

\subsection{High-efficiency high-gradient phase control}
\label{High-efficiency high-gradient phase control}
\begin{figure*}[htp]
\centering
\includegraphics[width=1\linewidth]{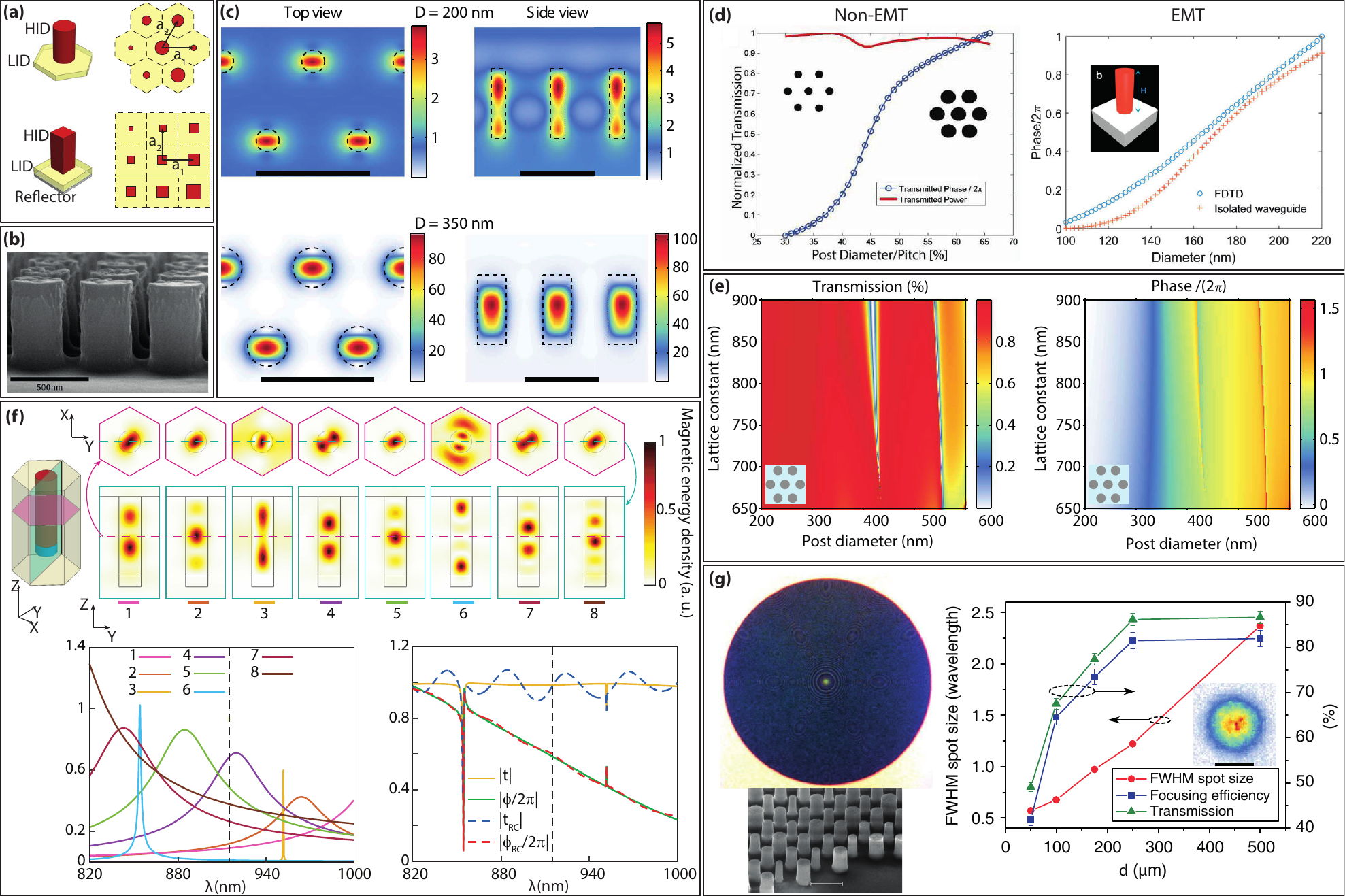}
\caption{\textbf{Operation principles of HCAs}. \textbf{(a)} Schematic illustration of some possible HCA configurations with different nano-post shapes and lattice structures. HID: high-index dielectric, LID: low -index dielectric. \textbf{(b)} Scanning electron micrograph of a graded index lens, etched directly into a silicon wafer~\cite{West2014OptExpress}.\textbf{(c)} Simulated magnetic energy density in a periodic array of $\alpha$-Si nano-posts plotted in cross sections perpendicular to (left), and passing through (right) the nano-posts' axes. For larger nano-posts, the field is highly confined inside the nano-posts. The scale bars are 1~$\mathrm{\mu}$m~\cite{Arbabi2015NatCommun}. \textbf{(d)} Simulated transmission phase for $\alpha$-Si nano-posts operating beyond the EMT regime (left,~\cite{Vo2014IEEEPhotonTechLett}), and TiO$_2$ nano-posts operating within the EMT regime (right,~\cite{Khorasaninejad2016NanoLett}). \textbf{(e)} Simulated transmission amplitude and phase of a periodic array of circular $\alpha$-Si nano-posts versus posts diameter and lattice constant~\cite{Arbabi2015NatCommun}. \textbf{(f)} Top: Magnetic energy distribution of optical resonances inside an $\alpha$-Si nano-post that contribute to the transmission amplitude and phase of the structure around 900~nm~\cite{Kamali2016NatCommun}. Bottom: Reconstruction of the nano-post transmission and phase using the frequency responses of the resonance modes. The left figure shows relative contributions of the different modes. \textbf{(g)} Near-IR lenses fabricated with $\alpha$-Si HCAs, and their performance as high-NA lenses with high efficiency~\cite{Arbabi2015NatCommun}. The lenses are designed to focus light emitted by a single-mode optical fiber to a diffraction-limited spot. The lenses have NAs from $\sim$0.5 to $\sim$0.97, and measured focusing efficiencies of 82$\%$ to 42$\%$. Scale bar: 1$\mathrm{\mu}m$.}
\label{Fig2_PhaseControl}
\end{figure*}

Since high-contrast transmitarrays (HCA) are central to the most of the discussions of this section, we first briefly discuss their operation. We are primarily interested in the two-dimensional HCAs, and therefore we consider their case here, although much of the discussions are also valid for the one-dimensional case. In general, these devices are based on high-refractive-index dielectric nano-scatterers surrounded by low-index media~\cite{Lalanne1998OptLett,Lalanne1999JOSAA,Fattal2011AdvPhoton,ArbabiCLEO2014,
Arbabi2015NatCommun,Arbabi2015NatNano,Kruk2016APLPhotonics,
Khorasaninejad2016Science,Paniagua2018NanoLett}. The structure can be symmetric (i.e., with the substrate and capping layers having the same refractive indexes)~\cite{Kamali2016LaserPhotonRev} or asymmetric~\cite{Lalanne1998OptLett,Lalanne1999JOSAA,Vo2014IEEEPhotonTechLett,Arbabi2015NatCommun}. Depending on the materials and the required phase coverage, the thickness of the high-index layer is usually between 0.5$\lambda_0$ and $\lambda_0$, where $\lambda_0$ is the free space wavelength. Typically, these structures are designed to be compatible with conventional microfabrication techniques; therefore, they are composed of nano-scatterers made from the same material structure and with the same thickness over the device area. Various scatterers can have different cross-sections in the plane of the metasurface, but the cross-section of any single scatterer is usually kept the same along the layer thickness to facilitate its fabrication using binary lithography techniques. For wavefront shaping, the nano-scatterers should be on the vertices of a subwavelength lattice that satisfies the Nyquist sampling criterion~\cite{Kamali2016LaserPhotonRev} in order to avoid excitation of unwanted diffraction orders. For simplicity, the lattices are usually selected to be periodic. Figure~\ref{Fig2_PhaseControl}a shows two typically used structures with triangular and square lattices~\cite{Fattal2011AdvPhoton,Lalanne1999JOSAA}. For polarization-independent operation (in the case of normal incidence with small deflection angle) the nano-scatterers should have symmetric cross-sections such as circles, squares, regular hexagons, etc. (Fig.~\ref{Fig2_PhaseControl}a). Similar to high-contrast gratings~\cite{Fattal2010NatPhoton,Lu2010OptExp,Klemm2013OptLett}, these structures can also be used in reflection mode by backing them with a metallic or dielectric reflector~\cite{Yang2014NanoLett,Arbabi2017Optica,Arbabi2017NatPhoton} [Fig.~\ref{Fig2_PhaseControl}a, bottom], or by properly selecting their thicknesses~\cite{ Arbabi2014CLEO_Retro,Hong2017ApplOpt}.

The first HCA diffractive devices demonstrated by Lalanne et al. (referred to as blazed binary diffractive devices at the time) were designed to operate in an effective medium theory (EMT) regime where only one transverse mode could be excited in the HCA layer~\cite{Lalanne1998OptLett,Lalanne1999JOSAA}. In 2011, it was suggested by Fattal et al.~\cite{Fattal2011AdvPhoton}, and later demonstrated~\cite{ArbabiCLEO2014,Vo2014IEEEPhotonTechLett,Arbabi2015NatCommun} that using higher index materials (Si or $\alpha$-Si instead of TiO$_2$) can result in devices with higher efficiency for large deflection angles, despite a departure from the EMT regime (where the lattice constant is larger than the structural cut-off~\cite{Lalanne1999JOSAA}, yet it is small enough to avoid unwanted diffraction~\cite{ArbabiCLEO2014,Vo2014IEEEPhotonTechLett}). It is worth noting that even for the lower index materials like TiO$_2$, the optimal operation regime seems to be where the lattice is designed just below the structural cut-off~\cite{Lalanne1999JOSAA_Multimode}. It is worth noting that depending on the design parameters, higher index devices (such as silicon ones) can operate in the EMT regime~\cite{West2014OptExpress}. One example of such devices is shown in Fig.~\ref{Fig2_PhaseControl}b, where a graded index lens was etched into a silicon wafer to focus light inside the wafer~\cite{West2014OptExpress}. The ability of the EMT blazed binary structures to significantly outperform the conventional \'{e}chelette gratings is dominantly attributed to the waveguiding effect of the nano-posts that results in a sampling of the incoming and outgoing waves with small coupling between adjacent nano-posts~\cite{Lalanne1999JOSAA_Multimode}. In devices using higher refractive index materials like silicon, the coupling between adjacent nano-posts remains small even above the structural cut-off. In Fig.~\ref{Fig2_PhaseControl}c, the simulated magnetic energy density is plotted for $\alpha$-Si nano-posts, showing that the field is highly confined inside the silicon nano-posts which reduces the coupling between nanoposts. In this case, there are multiple propagating transverse modes inside the layer~\cite{Arbabi2015NatCommun}. In addition, due to the larger number of resonances, the transmission phase for the nano-posts with the higher refractive index is a steeper function of the nano-post's size (compared to the EMT structures) as shown in Fig.~\ref{Fig2_PhaseControl}d~\cite{Vo2014IEEEPhotonTechLett,Khorasaninejad2016NanoLett}. This relieves the requirements of the nano-posts aspect ratio and makes their fabrication more feasible~\cite{Vo2014IEEEPhotonTechLett}.

A second effect of the high field confinement is that the behavior of the structure becomes more insensitive to the lattice parameters, as shown in Fig.~\ref{Fig2_PhaseControl}e~\cite{Arbabi2015NatCommun}. More importantly, this means that the transmission phase of a nano-post is largely insensitive to its neighboring posts; therefore, adjacent nano-posts can have significantly different sizes without much degradation of their performance. This is in contrast to the dielectric Huygens metasurfaces~\cite{Staude2013ACSNano,Yang2014NanoLett,Chong2015NanoLett,Shalaev2015NanoLett,
Yu2015LaserPhotonRev,Decker2015AdvOptMat} where the coupling between neighboring scatterers results in significant performance degradation if the size of the neighboring scatterers changes too abruptly. In addition, unlike the Huygens metasurfaces, the high transmission amplitude and full 2$\pi$-phase-coverage of the HCAs result from the contributions of multiple resonances. Such resonances are shown for a typical $\alpha$-Si nano-post in Fig.~\ref{Fig2_PhaseControl}f~\cite{Kamali2016NatCommun}. An expansion of the optical scattering of the nano-posts to electric and magnetic multipoles is also possible~\cite{Kruk2016APLPhotonics}. However, capturing the full physics requires the use of higher order multipoles, and the expansion does not give much direct information about the contribution of each resonance to each of the multipole terms, or how they can be tailored for a specific application.

In recent years, multiple groups have demonstrated high-efficiency high-NA lenses using the HCA platform~\cite{Arbabi2015NatCommun,Vo2014IEEEPhotonTechLett,Zhan2016ACSPhotonics,
Khorasaninejad2016NanoLett,Zhou2017ACSPhotonics,Paniagua2018NanoLett}. Figure~\ref{Fig2_PhaseControl}g, shows one of the early demonstrations where lenses with NAs ranging from $\sim$0.5 to above 0.95 were demonstrated, with measured absolute focusing efficiencies from 82$\%$ to 42$\%$ depending on the NA, while keeping a close to diffraction limited spot.

In both EMT and non-EMT regimes, the standard design method for optical phase masks (lenses in particular) has been to extract the transmission (reflection) coefficient for a periodic array of nano-posts and use them directly to design aperiodic devices that manipulate the phase profile~\cite{Lalanne1999JOSAA,Fattal2011AdvPhoton,Vo2014IEEEPhotonTechLett,ArbabiCLEO2014,Arbabi2015NatCommun}. This design process is based on the assumptions that the sampling is local, there is not much coupling between the nano-posts, and the transmission phase and amplitude remain the same for different scattering angles. The validity of these assumptions starts to break at large deflection angles and contribute to the lower efficiency of the devices at such angles. More recently, a few methods have been proposed and demonstrated potential for increasing the efficiency of these devices~\cite{Byrnes2016OptExp,Arbabi2017SPIEPW,Sell2017NanoLett}. While periodic devices (i.e., blazed gratings) with measured efficiencies as high as 75$\%$ at 75-degree deflection angles have been demonstrated, the case for non-periodic devices is more challenging, and to the best of our knowledge, the absolute measured focusing efficiencies for lenses with NAs about 0.8 have been limited to slightly above 75$\%$~\cite{Arbabi2017SPIEPW}. Proper measurement and reporting of the efficiency is a very important parameter in phase control devices with high-gradients. A proper definition of efficiency for lenses is the power of light focused to a small area around the focal point (for instance a disk with a diameter that is two to three times the diffraction limited Airy diameter). With this definition, it is essential in experiments that a pinhole be used around the focal spot to block the light outside this area; otherwise, the measured value would be the transmission efficiency. It is also important to identify the illuminating beam size when measuring efficiencies. Using a beam smaller than the clear aperture of the lens will effectively reduce the NA of the lens and lead to an overestimation of the device efficiency. The type of power detector used may also significantly bias the efficiency measurements. The light focused by a high NA device has a wide angular spectrum and many detectors are sensitive to the incident angle of light. Ideally, a detector with a wide acceptance angle such as an integrating sphere should be used in the efficiency measurements.

\subsection{Simultaneous polarization and phase control}
In refractive optics, polarization and phase control are generally performed with different types of devices. Diffractive optical elements, on the other hand, have the ability to simultaneously control polarization and phase. For instance, polarization-gratings have been fabricated and used to deflect light based on its state of polarization~\cite{Nikolova1984OpticaActa,Tervo2000OptLett,Oh2008OptLett}. Spatially varying polarization control (along with the associated geometric phase) was demonstrated about two decades ago with computer-generated holograms~\cite{Bomzon2001OptLettC,Bomzon2001OptLettP,Hasman2002OptCommun,Biener2002OptLett,
Bomzon2002OptLett,Hasman2003ApplPhysLett,Chia-Ho2004SPIE,Levy2004OptLett}. For example, using the geometric phase (that changes sign with changing the incident light helicity), Hasman et al. demonstrated polarization dependent focusing, where the lens demonstrates positive optical power for one incident helicity, and an equal but negative optical power for the other one~\cite{Hasman2003ApplPhysLett}. Dielectric metasurfaces that used geometric phase for beam shaping were also introduced around the same time~\cite{Bomzon2002OptLett,Biener2002OptLett}.

More recently, metallic and dielectric metasurfaces have been widely used to control the polarization of light~\cite{Zhao2011PhysRevB,Yu2011Science,Mutlu2012OptExp,Yu2012NanoLett,Huang2012NanoLett,Cheng2013ApplPhysA,Pfeiffer2013ApplPhysLett,Huang2013NatCommun,Ni2013NatCommun,Chen2014JApplPhys,Ma2014OptMatExp,Lin2014Science,Liu2015OptLett,Zhang2015IEEEAntenn,Gao2015IEEETAP,Shi2015IEEEALP,Arbabi2015SPIEPW,Arbabi2015NatNano,Zheng2015NatNano,Backlund2016NatPhoton,Kruk2016APLPhotonics,Chen2016Nanotechnology,Wu2017NanoLett}. Many of these devices are uniform waveplates that simply transform the polarization of the input wave upon reflection or transmission~\cite{Zhao2011PhysRevB,Mutlu2012OptExp,Cheng2013ApplPhysA,Chen2014JApplPhys,Ma2014OptMatExp,Liu2015OptLett,Zhang2015IEEEAntenn,Gao2015IEEETAP}. Many of the other designs are based on polarization dependent scatterers that locally act as waveplates (usually half-waveplates), but have varying rotations across the metasurface~\cite{Hasman2003ApplPhysLett,Yu2011Science,Yu2012NanoLett,Huang2012NanoLett,Huang2013NatCommun,Ni2013NatCommun,Lin2014Science,Shi2015IEEEALP,Arbabi2015SPIEPW,Zheng2015NatNano,Backlund2016NatPhoton,Kruk2016APLPhotonics,Chen2016Nanotechnology,Wu2017NanoLett}. In most cases of this category, the polarization conversion actually comes as an unwanted corollary of using the geometric phase for wavefront manipulation~\cite{Hasman2003ApplPhysLett,Yu2011Science,Huang2012NanoLett,Huang2013NatCommun,Ni2013NatCommun,Lin2014Science,Zheng2015NatNano}. In other cases, the polarization dependent response of plasmonic and dielectric nano-scatterers has been utilized to independently control the phase under two orthogonal (usually linear) polarizations~\cite{Farmahini2013OptLett,Pors2013SciRep,Chen2014NanoLett,Ma2015SciRep,Chen2016Nanotechnology}.

\begin{figure*}[htp]
\centering
\includegraphics[width=1\linewidth]{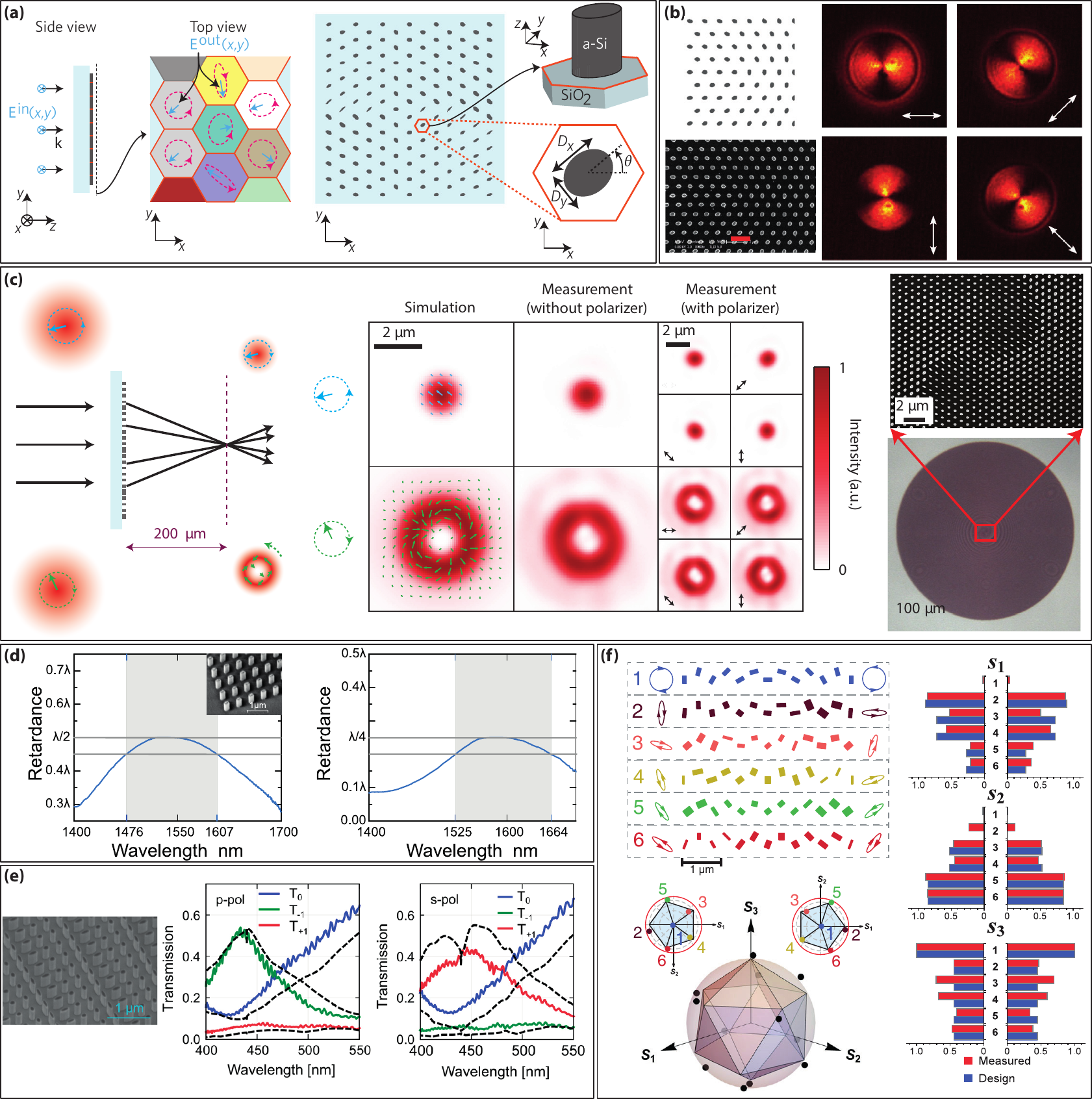}
\caption{\textbf{Simultaneous polarization and phase control with HCAs}. \textbf{(a)} Schematic illustration of the birefringent HCA nano-post with the ability to completely and independently control the phase and polarization of light~\cite{Arbabi2015NatNano}. \textbf{(b)} A spatially varying waveplate with the ability to convert horizontally and vertically-polarized input light to radially and azimuthally-polarized output light in the telecommunications band~\cite{Arbabi2015SPIEPW}. \textbf{(c)} An example of the possible devices that can be fabricated with the birefringent HCA structure. The device focuses the incident light to a diffraction-limited spot or a doughnut-shaped spot depending on its helicity. The same type of functionality can be achieved for any arbitrary orthogonal polarization basis (i.e., linear or elliptical)~\cite{Arbabi2015NatNano}. \textbf{(d)} Broadband half (left) and quarter (right) waveplates designed with birefringent HCAs. The waveplates have $\lambda$/20 bandwidths of $\sim$9$\%$~\cite{Kruk2016APLPhotonics}. \textbf{(e)} Linear polarization beam-splitters fabricated in GaN for visible~\cite{Emani2017ApplPhysLett}. \textbf{(f)} Several elliptical polarization beam-splitters fabricated in TiO$_2$ for green  light~\cite{Mueller2017PhysRevLett}.}
\label{Fig3_PolarizationPhaseControl}
\end{figure*}

In 2013, Pfeiffer and Grbic proposed a cascaded multilayer plasmonic metasurface with the ability to perform as a waveplate with independent control over the phase of \textit{x}- and \textit{y}-polarized transmitted light~\cite{Pfeiffer2013ApplPhysLett}. Although the achieved phase coverage was not complete, they were still able to design a lens composed of different quarter-waveplates that simultaneously changed the polarization from linear to circular and focused light~\cite{Pfeiffer2013ApplPhysLett}. Later, Yang et al. used a similar concept with a Si reflectarray to demonstrate half-waveplates with controllable \textit{x}- and \textit{y}-polarized reflection phases~\cite{Yang2014NanoLett}. Using the half-waveplates they demonstrated a beam deflector and vortex beam generator that simultaneously rotated the polarization by 90$^\circ$. In 2014, Arbabi et al. reported a generalization of this concept to achieve independent and simultaneous control of phase and polarization of light~\cite{Arbabi2017SciRep,Arbabi2015NatNano}. The idea is based on birefringent elements with the ability to completely and independently control the phase of two orthogonal linear polarizations, and freely choose the orientation of those directions (i.e., the optical axis directions). A schematic of a device with this ability is shown in Fig.~\ref{Fig3_PolarizationPhaseControl}a, where birefringent HCA nano-posts are used to provide this control~\cite{Arbabi2015NatNano}. They showed that with these two simple conditions, a metasurface can completely control the optical phase and polarization, with two different manifestations: first, an incident light with \textit{any} given polarization and phase can be transformed into an output light with \textit{any} desired polarization and phase. Second, given two orthogonal input polarizations (linear, circular, or elliptical), their phases can be independently controlled to perform two independent functions for the two polarizations. The second application comes at the small cost that the output polarization for each input will have the same polarization ellipse as the input one, but with the opposite handedness.

The birefringent HCA platform is especially suited for these applications as it provides the complete and independent control for orthogonal linear polarizations (with the proper choice of parameters). In addition, the weak coupling between adjacent nano-posts allows one to choose the dimensions and the orientation of each nano-post independently. This means that the control can be independently done at each point on a subwavelength lattice and with high efficiency~\cite{Arbabi2015NatNano}. A simple application for this platform is the implementation of spatially varying waveplates~\cite{Arbabi2015SPIEPW,Kruk2016APLPhotonics,Backlund2016NatPhoton}. For instance, using the device shown in Fig.~\ref{Fig3_PolarizationPhaseControl}b that is formed from half-waveplates with different rotation angles, the authors demonstrated a high-performance linear to azimuthal and radial polarization converter~\cite{Arbabi2015SPIEPW}. This platform can also be designed to achieve the same type of polarization conversion, while focusing light at the same time~\cite{Arbabi2015NatNano}. Another example of applications of this platform is shown in Fig.~\ref{Fig3_PolarizationPhaseControl}c, where a metasurface is designed to focus circularly polarized light either to a diffraction-limited spot or to a doughnut-shaped focus depending on the input handedness~\cite{Arbabi2015NatNano}. Generally, any two arbitrary functionalities (beam deflection, focusing, hologram projection, vortex beam generation, etc.) can be encoded into the two orthogonal polarizations. Various types of polarization-switchable holograms~\cite{Arbabi2015NatNano,Mueller2017PhysRevLett} and mode generators~\cite{Arbabi2015NatNano,Kruk2017arXiv} have also been demonstrated.

An important property of this platform is that the wide range of available design parameters (material system, high-index layer thickness, lattice constant, shape and size of the nano-posts) allows one to design waveplates with relatively wide bandwidths ($\sim$10$\%$) and low sensitivity to fabrication errors, as different groups have demonstrated in the telecommunication~\cite{Kruk2016APLPhotonics} and visible red~\cite{Backlund2016NatPhoton} bands. Figure~\ref{Fig3_PolarizationPhaseControl}d shows the retardance for two such designs that demonstrate a $\sim$9$\%$ bandwidth for quarter and half-waveplates.

Similar to the polarization-insensitive case, this concept and platform can be transformed to any electromagnetic band using a proper selection of material system and scaling. Recently different groups have demonstrated polarization beamsplitters for linear~\cite{Emani2017ApplPhysLett} and elliptical~\cite{Mueller2017PhysRevLett} polarizations using GaN and TiO$_2$, respectively. The measurement results of these devices are shown in Figs.~\ref{Fig3_PolarizationPhaseControl}e~\cite{Emani2017ApplPhysLett} and~\ref{Fig3_PolarizationPhaseControl}f~\cite{Mueller2017PhysRevLett}, respectively. As seen from Fig.~\ref{Fig3_PolarizationPhaseControl}e, the efficiency of these devices operating at visible wavelengths are still limited to about 50$\%$ (the efficiencies of the TiO$_2$ devices have not been reported), compared to the near-IR devices where efficiencies above 90$\%$ have been  achieved~\cite{Arbabi2015NatNano}. One possible cause can be the lower available refractive indexes in the visible that does not allow for full phase and polarization control with low coupling between nano-posts. The larger sensitivity to fabrication errors might be another reason for the lower efficiencies in the visible. In addition, the geometrical-phase wavefront manipulators that use waveplates (usually half-waveplates) with different rotations to impose a specific phase profile on a specific input polarization~\cite{Khorasaninejad2016Science,Wang2016NanoLett,Chen2017NanoLett} can now be thought of as a particular case of this generalized platform for phase and polarization control. Since for those devices only one type of nano-post (i.e., a single half-waveplate) is required, the dimensions of that nano-post can be designed such that the final device has higher efficiency and reduced sensitivity to fabrication errors or small changes in operation wavelength~\cite{Backlund2016NatPhoton,Kruk2016APLPhotonics}. However, similar to other geometric phase based devices, these optical elements only work for one polarization and the power in the other polarization is lost into unwanted diffraction, limiting their theoretical unpolarized-light efficiency to 50$\%$.

\subsection{Controlling chromatic dispersion}
\label{Controlling chromatic dispersion}
The bandwidth of operation is an important parameter for many types of optical devices. The definition of bandwidth depends on the actual functionality and the application of the device. In this section, we review the recent advancements in addressing the issue of chromatic dispersion in metasurfaces that manipulate the phase profile of light, i.e., the change in the performance of such devices when the wavelength is changed. Examples of such devices include beam deflectors, lenses, holograms, etc. where the deflection angle, focal distance, and size of the holographic image (or the reconstruction distance) change with changing the wavelength, respectively~\cite{Born1999,O'Shea2004}. This is a fundamentally different issue from the wavelength dependence of some other types of metasurfaces such as waveplates or absorbers that do not shape the beam. In addition, although related in some sense, this is physically different from the challenges faced in designing broadband (in the sense of having high efficiency over a broad bandwidth) dispersive gratings that show the regular dispersion characteristics of diffractive gratings.

\begin{figure*}[htp]
\centering
\includegraphics[width=1\linewidth]{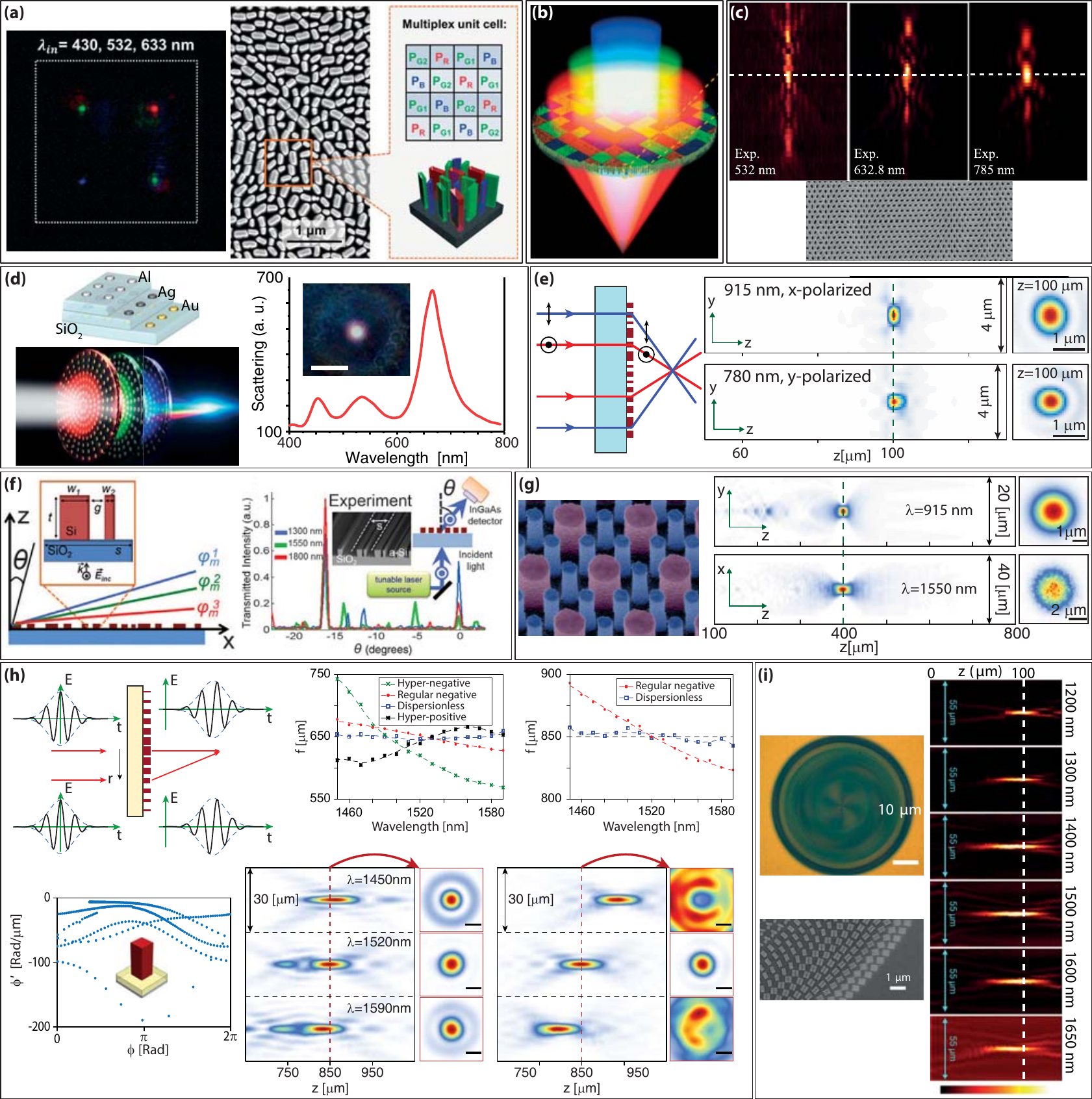}
\caption{\textbf{Multi-wavelength and dispersion-engineered metasurfaces}. \textbf{(a)} An example of a spatially-multiplexed multi-wavelength metasurface through interleaving of GaN meta-atoms. Four lenses are multiplexed to separate RGB colors~\cite{Chen2017NanoLett}. \textbf{(b)} Concept of a multi-wavelength metasurface lens multiplexed through segmentation~\cite{Jun2017JourOpt}. \textbf{(c)} Multi-wavelength metasurface lens designed through holographic superposition of fields~\cite{Zhao2015SciRep}. \textbf{(d)} Multi-wavelength Fresnel-zone-plate lens designed using frequency-dependent scatterers~\cite{Avayu2017NatCommun}. \textbf{(e)} Double-wavelength metasurfaces designed with birefringent meta-atoms~\cite{Arbabi2016OptExp}. \textbf{(f)} Multi-wavelength metasurface grating designed with mutiscatterer unit-cells (meta-molecules)~\cite{Aieta2015Science}. \textbf{(g)} Polarization insensitive multi-wavelength lenses designed with dielectric meta-molecules~\cite{Arbabi2016Optica}. \textbf{(h)} Dispersion-engineered metasurfaces designed using meta-atoms with independent control of phase and group delays. Focusing mirrors and gratings with inverse, zero, and increased dispersion are demonstrated~\cite{Arbabi2017Optica}. \textbf{(i)} Achromatic lenses and gratings designed using plasmonic meta-atoms with varying phases and phase derivatives~\cite{Wang2017NatCommun}.}
\label{Fig4_Chromatic}
\end{figure*}

The issue of chromatic dispersion in diffractive optics has been studied and is well-known for a long time~\cite{Born1999,O'Shea2004,Miyamoto1961JOSA,Faklis1995ApplOpt}. Unlike refractive optics where chromatic dispersion is caused by material dispersion, the chromatic dispersion of diffractive optical devices results from the structural dispersion~\cite{Miyamoto1961JOSA,Faklis1995ApplOpt}. The most well-known case of this behavior is the diffraction grating where the angle of deflection for each order and the rate of change of that angle with wavelength is solely determined by the pitch of the grating. In diffractive lenses, this property is manifested through a significant change in the focal distance with changing the wavelength, thus significantly limiting the applications of diffractive-only lens systems. To address this issue to some extent, multi-order multiwavelength diffractive elements were used~\cite{Faklis1995ApplOpt}. Such devices are in essence similar to multi-order gratings where different orders of interest are blazed to have high efficiencies at different wavelengths, where the deflection angles (or focal distances) of the different orders are the same. Recently there has been a great deal of interest in addressing the chromatic dispersion issue in metasurfaces~\cite{Eisenbach2015OptExp,Aieta2015Science,Cheng2015JOSAB,Zhao2015SciRep,Khorasaninejad2015NanoLett,
Arbabi2016Optica,Arbabi2016OptExp,Arbabi2016SciRep,Hu2016ACSNano,Wang2016NanoLett,Wang2016SciRep,
Avayu2017NatCommun,Chen2017NanoLett,Jun2017JourOpt,Sell2017AdvOptMat,Arbabi2016CLEO_Displess,
Arbabi2017Optica,Khorasaninejad2017NanoLett,Wang2017NatCommun}. Most of the demonstrated methods result in \textit{multi-wavelength} devices and are in principle similar to the multi-order diffractive optical elements~\cite{Eisenbach2015OptExp,Aieta2015Science,Zhao2015SciRep,Khorasaninejad2015NanoLett,
Arbabi2016Optica,Arbabi2016OptExp,Arbabi2016SciRep,Wang2016NanoLett,Avayu2017NatCommun,
Chen2017NanoLett,Jun2017JourOpt,Sell2017AdvOptMat}. More recently, a method based on independent control of phase and its wavelength derivative (dispersion) has been introduced~\cite{Arbabi2016CLEO_Displess} and used to implement achromatic~\cite{Arbabi2016CLEO_Displess,Arbabi2017Optica,Khorasaninejad2017NanoLett,Wang2017NatCommun,Chen2017NanoLett} and dispersion-engineered~\cite{Arbabi2017Optica} metasurface diffractive devices. Here we first review the multi-wavelength devices and discuss several methods that have been utilized to demonstrate them. Then we explain the dispersion-phase control and mention the devices that have been demonstrated using this method.

The simplest method used to demonstrate mutiwavelength devices is based on spatial multiplexing of metasurfaces~\cite{Arbabi2016SciRep,Zhao2016OptLett,Hu2016ACSNano,Lin2016NanoLett,Wang2016NanoLett,
Jun2017JourOpt,Chen2017NanoLett}.   In this method, multiple metasurfaces are designed for multiple wavelengths (one for each wavelength). For simplicity of fabrication, all metasurfaces are typically designed with the same material and with the same thickness. Two slightly different approaches can then be taken to combine such multi-wavelength devices: first, the meta-atoms can be interleaved on a wavelength scale scheme~\cite{Arbabi2016SciRep,Wang2016NanoLett,Chen2017NanoLett}. One example of such devices is shown in Fig.~\ref{Fig4_Chromatic}a, where four different lenses (two for green, one for red, and one for blue) are interleaved to focus the colors to different points~\cite{Chen2017NanoLett}. The second approach is based on dividing the device aperture to macroscopic segments and assigning different wavelengths to different segments. An example of such a device is schematically shown in Fig.~\ref{Fig4_Chromatic}b, where a lens is designed to have the same focal length at the red, blue, and green wavelengths~\cite{Jun2017JourOpt}. This method can also be applied to conventional diffractive elements. The second method of multi-wavelength design is combining the required phases for different wavelengths to the structure in a holographic design manner (i.e., by adding the transmission coefficients at different wavelengths)~\cite{Zhao2015SciRep}. A lens designed to focus three visible wavelengths to the same focal distance and its axial plane measurement results are shown in Fig.~\ref{Fig4_Chromatic}c. The main problem with all of the mentioned methods is that the multi-wavelength operations comes at the expense of efficiency and elevated background resulting from the imperfect phase.

Another method to achieve multi-wavelength operation is using frequency selective meta-atoms~\cite{Avayu2017NatCommun}. Similar methods were used in the microwave domain to demonstrate multi-band reflectarrays as early as the 1990s~\cite{Huang1991APS}. Figure~\ref{Fig4_Chromatic}d shows the schematics and measurement results for one such device that is based on three different plasmonic layers. Each plasmonic layer has a different resonance wavelength, resulting in the layer becoming non-transparent. The authors have used this effect to show multiple Fresnel zone plate (FZP) devices. Another (to some extent related method) is based on using birefringent meta-atoms to create different phase (or amplitude in the case of FZPs) profiles for two wavelengths under orthogonal polarizations~\cite{Eisenbach2015OptExp,Arbabi2016OptExp}. This is schematically shown in Fig.~\ref{Fig4_Chromatic}e, along with the measurement results of one such lens. Recently, Wang et al. have used a combination of the polarization dependent and spatial multiplexing methods to demonstrate a multi-color hologram~\cite{Wang2017Optica}.

A different design is based on using multiple meta-atoms per unit-cell (i.e., a meta-molecule) in order to have multiple phase control parameters. This method has been used to demonstrate various multiwavelength gratings~\cite{Aieta2015Science} and lenses~\cite{Khorasaninejad2015NanoLett,Arbabi2016Optica}. Figures~\ref{Fig4_Chromatic}f~\cite{Aieta2015Science} and \ref{Fig4_Chromatic}g~\cite{Arbabi2016Optica} show schematics and measurement results for such typical gratings and lenses. In the approach presented in~\cite{Arbabi2016Optica}, a set of meta-molecules are designed such that they can independently provide any combination of two phases at two different wavelengths. The concept is general and can be used to design efficient multiwavelength and multifunctional devices. Finally, one could use optimization methods (for periodic structures up to now) to demonstrate multiwavelength devices~\cite{Sell2017AdvOptMat}. All of these methods can be and have been utilized to design devices with the same or different functions at multiple wavelengths.

In 2016, Arbabi et al. introduced a method to address the issue of chromatic dispersion over a continuous bandwidth~\cite{Arbabi2016CLEO_Displess} and used it to show a focusing mirror (NA$\sim$0.3, $f=$850~$\mathrm{\mu}$m at 1520~nm) with significantly diminished dispersion over a 10$\%$ bandwidth. The method is based on meta-atoms with the ability to independently control the phase and group delays. To have achromatic behavior, portions of a pulse that hit different parts of a lens should arrive at the focus with the same phase and group delays. This is schematically shown in Fig.~\ref{Fig4_Chromatic}h.  Mathematically, it is simpler to use the equivalent terminology of phase-dispersion (i.e., phase and its wavelength derivative) to describe this behavior. One meta-atom comprising of an $\alpha$-Si nano-post on a low index material backed by a metallic mirror, and the coverage it provides in the phase-dispersion plane is shown in Fig.~\ref{Fig4_Chromatic}h~\cite{Arbabi2017Optica}. It is seen that for each phase, there are about four different available dispersion values. The idea was also generalized to devices that demonstrate non-zero dispersion. For instance, gratings and focusing mirrors with reverse (i.e., positive) and enhanced (hyper-negative) dispersion were also demonstrated~\cite{Arbabi2017Optica}. The measured focal distances and intensity distributions for a few focusing mirrors are shown in Fig.~\ref{Fig4_Chromatic}h. The devices were demonstrated to have focusing efficiencies of approximately 50$\%$ over the design bandwidth. Similar structures as in Ref.~\cite{Arbabi2016CLEO_Displess} but fabricated in TiO$_2$ were also used to demonstrate achromatic focusing mirrors in the visible with about 10$\%$ bandwidth, and <20$\%$ efficiency~\cite{Khorasaninejad2017NanoLett}.

Focusing mirrors based on the same phase-dispersion control concept, but using plasmonic scatterers were used to demonstrate reduced chromatic dispersion for focusing mirrors with a few different focal distances and NA values (NA$\sim$0.2-0.3, $f\sim$150~$\mathrm{\mu}$m-65~$\mathrm{\mu}$m)~\cite{Wang2017NatCommun}. The devices work under a single circular polarization, showed reduced chromatic dispersion from 1200~nm to 1650~nm, and have average efficiencies below 12$\%$. Recently, Chen et al. demonstrated a transmissive lens working in the visible and showing reduced dispersion from 470~nm to 670~nm~\cite{Chen2018NatNano}. The device has a 70-$\mathrm{\mu}$m focal distance and a numerical aperture of 0.2, and focuses one circular polarization with <20$\%$ efficiency. Wang et al. demonstrated transmissive lenses in the visible (400-660~nm range) with average efficiencies about 40$\%$ that show reduced chromatic dispersion~\cite{Wang2018NatNano}. The lenses have diameters of 60~$\mathrm{\mu}$m and numerical apertures smaller than 0.15.

A very important issue regarding metasurfaces with increased bandwidth is that a comparison based on the absolute or relative bandwidth ($\Delta\lambda$ or $\Delta\lambda/\lambda$) could be misleading. Therefore, it is important to have a proper criterion for comparing how much an achromatization technique can in fact increase the bandwidth of various devices with different wavelengths, NAs, and sizes. Since most of the work in the field is concerned with designing lenses and focusing mirrors, here we use a criterion that is relevant in this context. To reduce the imaging quality beyond acceptability, the focus should not move more than the depth of focus. In order to make this independent of wavelength and image sensor characteristics or target resolution, we use the criterion suggested in~\cite{Arbabi2016NatCommun} that uses the Rayleigh range of the focused beam instead of the depth of focus. This way, to degrade the performance below acceptability, the focus should move such that the spot area is increased to twice its optimal value. For a conventional diffractive lens (or a typical metasurface lens) working at a wavelength $\lambda$, the  resulting acceptable operation bandwidth is then given by $\Delta\lambda_0=(\pi/2\mathrm{ln(2)})\lambda^2/(f\mathrm{NA}^2)$~\cite{Arbabi2016NatCommun}. Therefore, to compare the operation bandwidths for different devices, one should normalize it to $\Delta\lambda_0$, indicating how much the method can actually \textit{increase} the bandwidth. Here we calculate this ratio for the five works discussed above~\cite{Arbabi2016CLEO_Displess,Arbabi2017Optica,Khorasaninejad2017NanoLett,
Wang2017NatCommun,Chen2018NatNano,Wang2018NatNano}. For the devices in~\cite{Arbabi2016CLEO_Displess,Arbabi2017Optica} and~\cite{Khorasaninejad2017NanoLett} this ratio is $\Delta\lambda_{ach}/\Delta\lambda_0\sim$1.9, for the mirrors in~\cite{Wang2017NatCommun} the maximum value of this ratio is $\Delta\lambda_{Ach}/\Delta\lambda_0\sim$0.7, and for the lenses demonstrated in~\cite{Chen2018NatNano} and~\cite{Wang2018NatNano} $\Delta\lambda_{Ach}/\Delta\lambda_0\sim$0.9 and 1.08, respectievely. Therefore, the best results have yet been demonstrated with dielectric focusing mirrors~\cite{Arbabi2016CLEO_Displess,Arbabi2017Optica,Khorasaninejad2017NanoLett}.

The formula of the operation bandwidth given above shows that it is significantly easier to design achromatic devices that have smaller physical apertures and also smaller numerical apertures~\cite{Cheng2015JOSAB}. This is directly related to the time delay between rays that propagate through the center of the lens and the circumference of the lens as shown in Fig.4h (top left). Equivalently, the maximum size and numerical aperture of a lens that can be designed with a specific platform are limited by the highest quality factors (Qs) of the meta-atoms that can be reliably achieved. Since for higher Qs it is harder to maintain a linear phase versus frequency relation, it becomes harder and harder to design large-aperture high-NA devices that operate over a wide bandwidth~\cite{Arbabi2017Optica}. In addition, the efficiency and achievable level of geometrical aberration correction are limited by how well the phase and dispersion can be independently controlled~\cite{Arbabi2017Optica}. The solution is to find meta-atoms that support multiple high Q resonances that can linearize the phase versus frequency relation over a wide range; a challenging task that has not been yet overcome. For simplicity, here we ignored the effect of efficiency that can change the actual bandwidth significantly by changing the signal to noise ratio. For a more detailed discussion of this issue see~\cite{Arbabi2016NatCommun,Engelberg2017OptExp}. Another route to overcome the chromatic dispersion issue might be computationally enhanced imaging using unconventional metasurface phase masks~\cite{Colburn2018SciAdv}.

\subsection{Angular response control with metasurfaces}

\begin{figure*}[htp]
\centering
\includegraphics[width=1\linewidth]{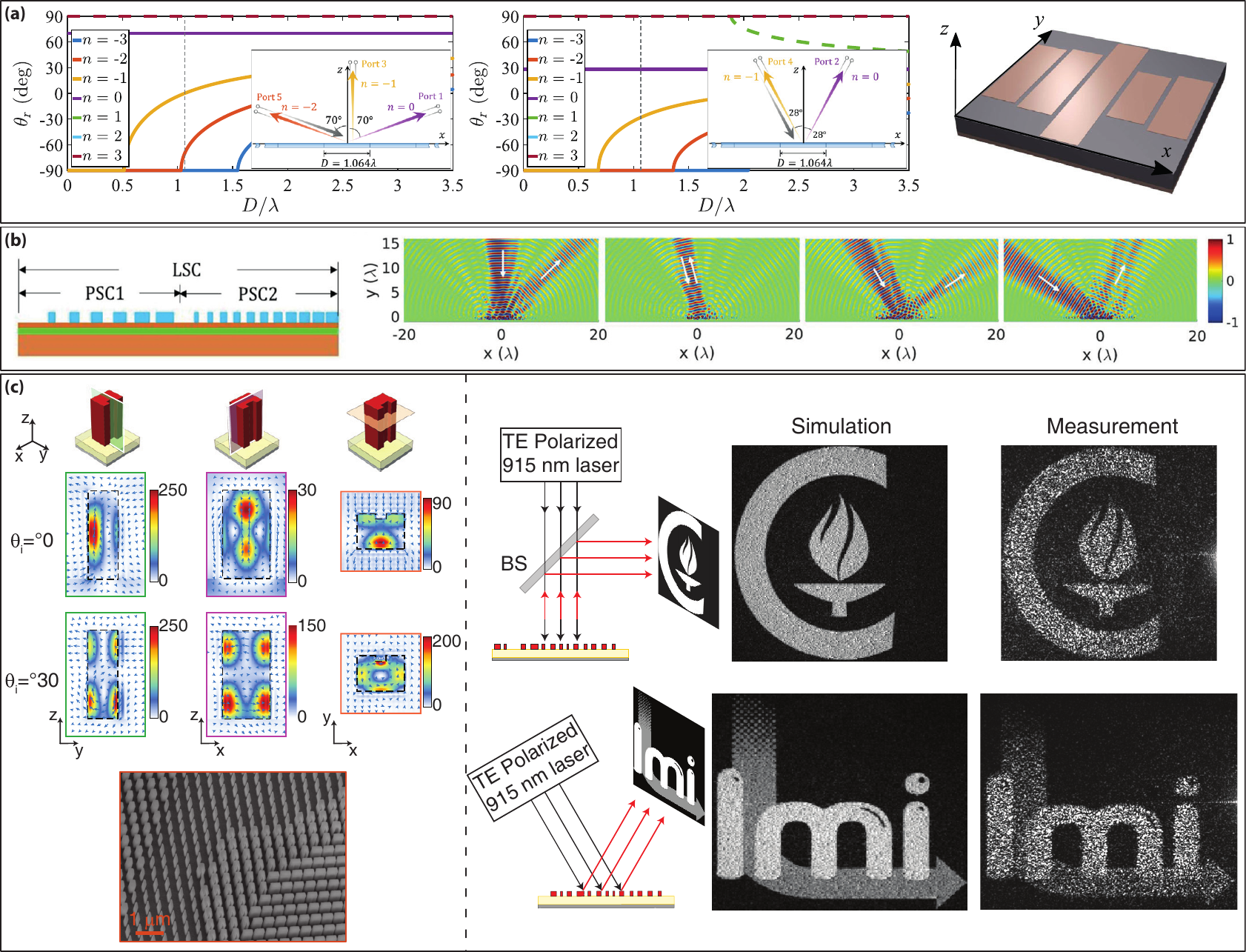}
\caption{\textbf{Metasurfaces with angular response control}. \textbf{(a)} Multichannel control of microwave metasurfaces through use of allowed and prohibited diffraction channels. Left and middle: A slightly super-wavelength Littrow grating (the period is shown with the dashed black line) that retroreflects under five different blaze angles~\cite{Asadchy2017PRX}. Right: schematic of the platform used to demonstrate multi-angle metasurfaces. \textbf{(b)} One-dimensional high-contrast dielectric reflective gratings numerically optimized to blaze different diffraction orders under illumination from different input angles~\cite{Cheng2017SciRep}. \textbf{(c)} High-contrast dielectric reflectarray designed to provide independent phase control under two different incident angles~\cite{Kamali2017PRX}. The U-shaped meta-atom allows for independent control of symmetric and anti-symmetric modes inside it, thus enabling devices with independent arbitrary functions at different incident angles. Gratings with different momenta, and holograms projecting different images were demonstrated.}
\label{Fig5_Angular}
\end{figure*}

In the past two sections, we reviewed the recent advances in independently controlling light based on the polarization and wavelength degrees of freedom of the input light. Another degree of freedom for the input wave is its spatial distribution (i.e., the spatial mode domain). The simplest basis for this degree of freedom is the plane-wave basis. In other words, metasurfaces with the ability to independently control light incoming from different angles are also of interest. Developing such metasurfaces is challenging mainly because of two reasons: first, due to their diffractive nature, metasurfaces show a high angular response correlation. The most typical case for this behavior is that of a diffraction grating. Defined only by their periodicity, gratings have certain diffraction orders (i.e., the Floquet-Bloch modes), each one with an associated grating momentum. The case is more complicated, but fundamentally similar for non-periodic devices like Fresnel lenses and holograms due to the existence of Fresnel zone boundaries~\cite{Miyamoto1961JOSA,Faklis1995ApplOpt,Arbabi2016Optica}.  The second property that makes it challenging to control the angular response with metasurfaces is that the scattering response of meta-atoms is generally insensitive to the incident angle. This can be thought of as a special case of the more general optical angular memory effect~\cite{Feng1988PhysRevLett}. In fact, recently it has been shown that metasurface scattering media have unusually large optical angular memory ranges~\cite{Jang2018NatPhoton}. In this section, we mention three recent works that have addressed this issue~\cite{Asadchy2017PRX,Cheng2017SciRep,Kamali2017PRX}. The first two~\cite{Asadchy2017PRX,Cheng2017SciRep} have focused on controlling the blaze profile for multi-order gratings, and the third has developed a platform that provides independent wavefront control under two incident angles, allowing for achieving two arbitrary functions from a single device illuminated at different angles~\cite{Kamali2017PRX}.

One could treat a multi-order diffraction grating as a multi-port system with a regular scattering matrix notation, in case one is only interested in illumination from a finite number of incident angles~\cite{Asadchy2017PRX}. The ports of interest are then determined by the angles of incidence, and their existing diffraction orders. One such case is shown in Fig.~\ref{Fig5_Angular}a left, where a slightly super-wavelength grating (a grating whose period is larger than the wavelength) with the zeroth and the first ($\pm$1) orders propagating under normal illumination is assumed. For the specific periodicity shown by the vertical dashed black line, illumination from 28$^\circ$ will result in only the zeroth and -1 (to -28$^\circ$) orders [Fig.~\ref{Fig5_Angular}a, middle]. The authors have designed a microwave reflectarray that retroreflects under illumination from these specific angles using the platform schematically shown in Fig.~\ref{Fig5_Angular}a, right. The device is essentially working as a Littrow grating with -2 and -1 orders being blazed under illuminations from $\pm$70$^\circ$ and $\pm$28$^\circ$, respectively. Due to the limited angular dependence of the metallic reflective elements, high efficiency operation of the device depends on the fact that for each input channel there are only a very limited number of possible output channels (in this specific case only three ports for normal and $\pm$70$^\circ$, and two ports for $\pm$28$^\circ$). In addition, the number of conditions that should be satisfied can also be decreased using symmetries of the required operation. For instance, the required surface impedances for retroreflection under incident angles of $+\theta_i$ and $-\theta_i$ are in fact the same~\cite{Asadchy2017PRX}.

Cheng et al. used a high-contrast dielectric grating backed by a metallic mirror to independently control the diffraction angle for different incident angles under TM-polarized illumination~\cite{Cheng2017SciRep}. Since the high-contrast devices have a significant angular dependence for TM-polarized light~\cite{Kamali2016NatCommun}, this structure [Fig.~\ref{Fig5_Angular}b, left] allows for better independent control of the blaze profile. Using this structure and an optimization method, they demonstrated a few different devices with the ability to independently control the blaze profile for a few angles. The response of one such device under four different incident angles is shown in Fig.~\ref{Fig5_Angular}b. The structure has a periodicity of about fourteen wavelengths with many possible diffraction orders, and therefore these simulation results confirm the ability of the device to control blazing profile based on the incident angle to some extent.

The main limitation of both methods mentioned above is that they are still limited by the existing diffraction orders of the grating, and they are only applicable to periodic structures. Recently Kamali et al. introduced a platform that allows for independent phase control under two different incident angles~\cite{Kamali2017PRX}. The structure is based on U-shaped high-index nano-posts backed by a metallic reflector, shown in Fig.~\ref{Fig5_Angular}c, left. The specific shape allows for independent control of reflection phases using the symmetric and anti-symmetric resonances inside the nano-posts. This platform allows for arbitrary independent control of the device function under different input angles. For instance, the authors demonstrated blazed gratings with different periodicities and a hologram that projects two different holographic images when illuminated from 0 and 30 degrees. The simulation and measurement results of the hologram are shown in Fig.~\ref{Fig5_Angular}c.

So far, all demonstrated platforms control the angular response of the metasurface for a discrete number of illumination angles. Controlling the behavior of a single-layer metasurface over a continuous range of angles is more challenging and requires elements that enable independent control of phase and its derivative with respect to angle~\cite{Kamali2017SPIEPW}.

\section{Applications of optical metasurfaces}

Metasurfaces have been utilized in different applications, ranging from emulating the conventional optical elements with miniaturized sizes, to providing new functionalities not feasible with conventional optical devices. In this section, we provide a brief overview of various applications of optical metasurfaces including wavefront manipulation, tunable and reconfigurable devices, conformal optics, and metasystems. 

\begin{figure*}[htp]
\centering
\includegraphics[width=2\columnwidth]{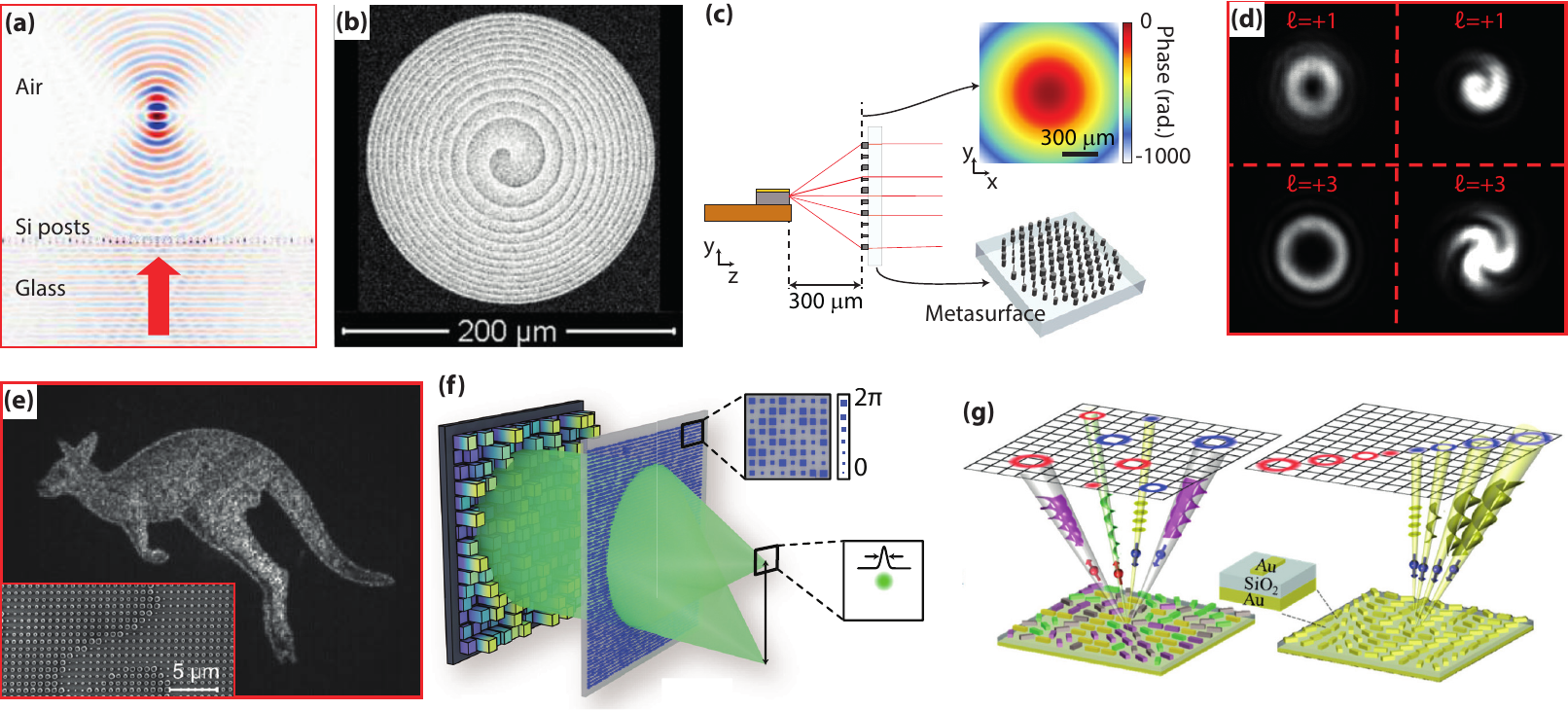}
\caption{\textbf{Wavefront shaping with phase control}. \textbf{(a)} A temporal snapshot of the simulated field distribution of an aspherical metasurface lens designed to shape the wavefront of light into a spherical phase profile. An input Gaussian beam is focused to a diffraction limited spot after passing through the polarization independent metasurface~\cite{Fattal2011AdvPhoton}. \textbf{(b)} A vortex-generating metasurface lens designed to focus the input beam with added orbital angular momentum (OAM). Scanning electron micrograph of the top view of the polarization independent vortex metasurface lens is shown~\cite{Vo2014IEEEPhotonTechLett}. \textbf{(c)} Schematic illustration of a metasurface collimating lens designed to collimate the output beam of a single-mode mid-infrared quantum-cascade laser~\cite{Arbabi2015OptExp}. \textbf{(d)} A metasurface OAM generator designed to generate blazed grating ``fork'' phase masks with different OAM orders. The measured intensity profiles (left), and the interferograms (right) of the OAM beams with different orders, which are generated through a metasurface under green light illumination~\cite{Ren2016SciRep}. \textbf{(e)} A metasurface designed to shape the wavefront of light to project a holographic image. Captured holographic image, created with a metasurface at the telecom band. (Inset) Scanning electron microscope image of a portion of the fabricated metasurface hologram~\cite{Wang2016Optica}. \textbf{(f)} A disorder-engineered metasurface designed to generate a random phase profile with wide angular scattering range. It allows for focusing light with a high numerical aperture over a wide field of view~\cite{Jang2018NatPhoton}. \textbf{(g)} Metasurfaces designed to generate multiplexed geometric phase profiles. Schematic illustration of two different methods for making shared-aperture metasurfaces: spatial multiplexing (left) and field superposition (right)~\cite{Maguid2016Science}.}
\label{fig:6_WavefrontShaping}
\end{figure*}

\begin{figure*}[htp]
\centering
\includegraphics[width=2\columnwidth]{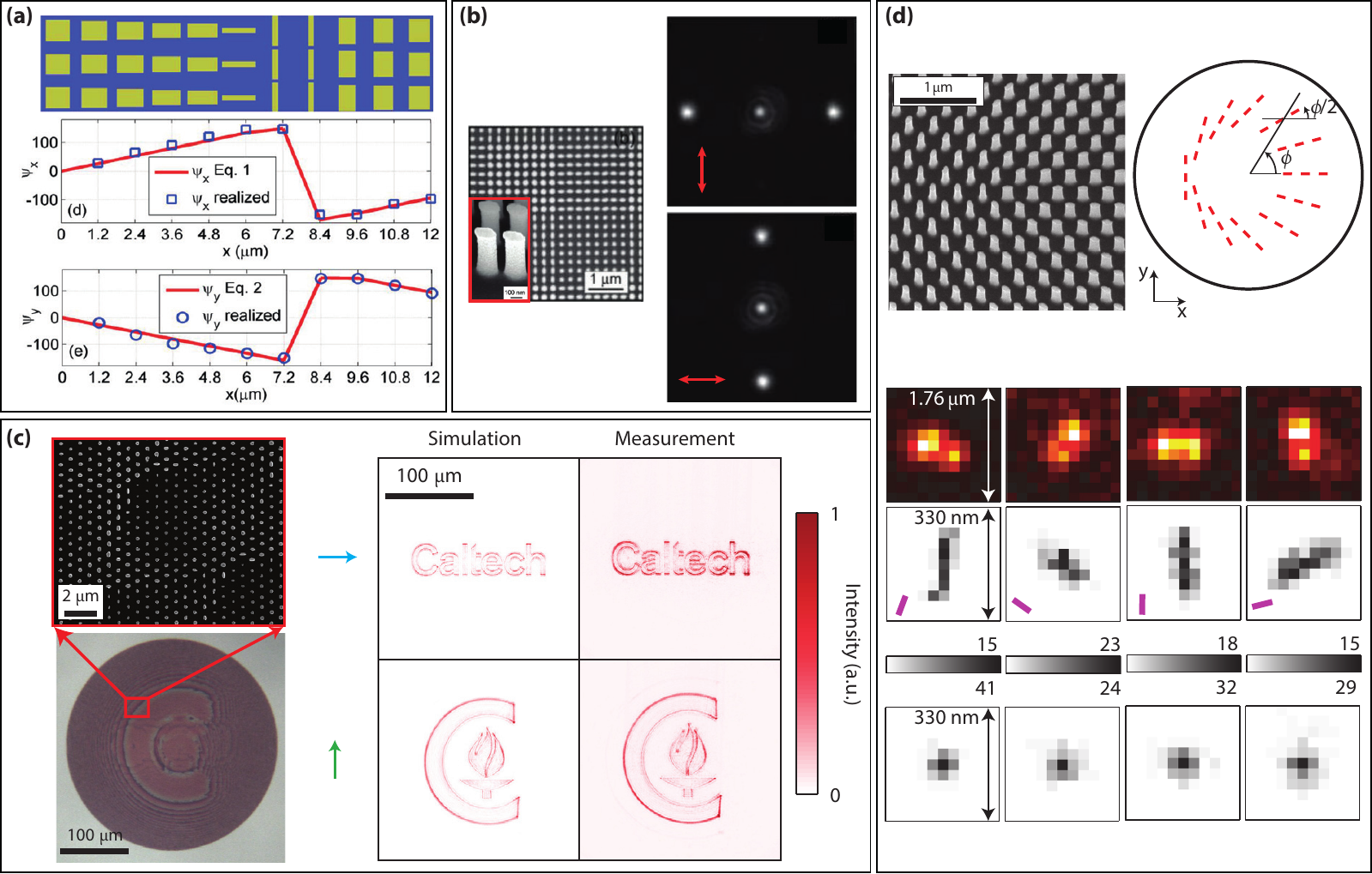}
\caption{\textbf{Wavefront shaping with polarization and phase control}. \textbf{(a)} A reflective grating designed to deflect \textit{x}- and \textit{y}-polarized light to different deflection angles. Schematic of one period of the device is shown on the top. Desired and simulated phase profiles at an operation wavelength of 8.06 $\mathrm{\mu}$m under x and y polarizations are shown on the bottom respectively \cite{Farmahini2013OptLett}. \textbf{(b)} A 2D grating with the ability to diffract \textit{x}-polarized light vertically and \textit{y}-polarized light horizontally is demonstrated. Scanning electron micrograph of the fabricated grating is shown on the left. Far-field intensities of the grating under \textit{x}- and \textit{y}-polarized light are shown on the right. The grating is measured under 976-nm illumination~\cite{Schonbrun2011NanoLett}. \textbf{(c)} A polarization switchable hologram generates two different images under horizontal and vertical polarizations~\cite{Arbabi2015NatNano}. Optical (bottom) and scanning electron micrograph (top) of the device are shown on the left. Simulated and measured intensity profiles under different linear polarizations are shown on the right. \textbf{(d)} Scanning electron micrograph of a portion of the device that converts azimuthally polarized light into \textit{y}-polarized light is shown on the left. The metasurface functions as spatially varying half-wave plates, with the principal axis orientation indicated by the red dashed lines (top, right). Images of four molecules (top row), and apparent displacement of those molecules through a z-scan with (bottom row) and without (middle row) the metasurface mask~\cite{Backlund2016NatPhoton}.}
\label{fig:7_PolPhase_application}
\end{figure*}

\begin{figure*}[htp]
\centering
\includegraphics[width=2\columnwidth]{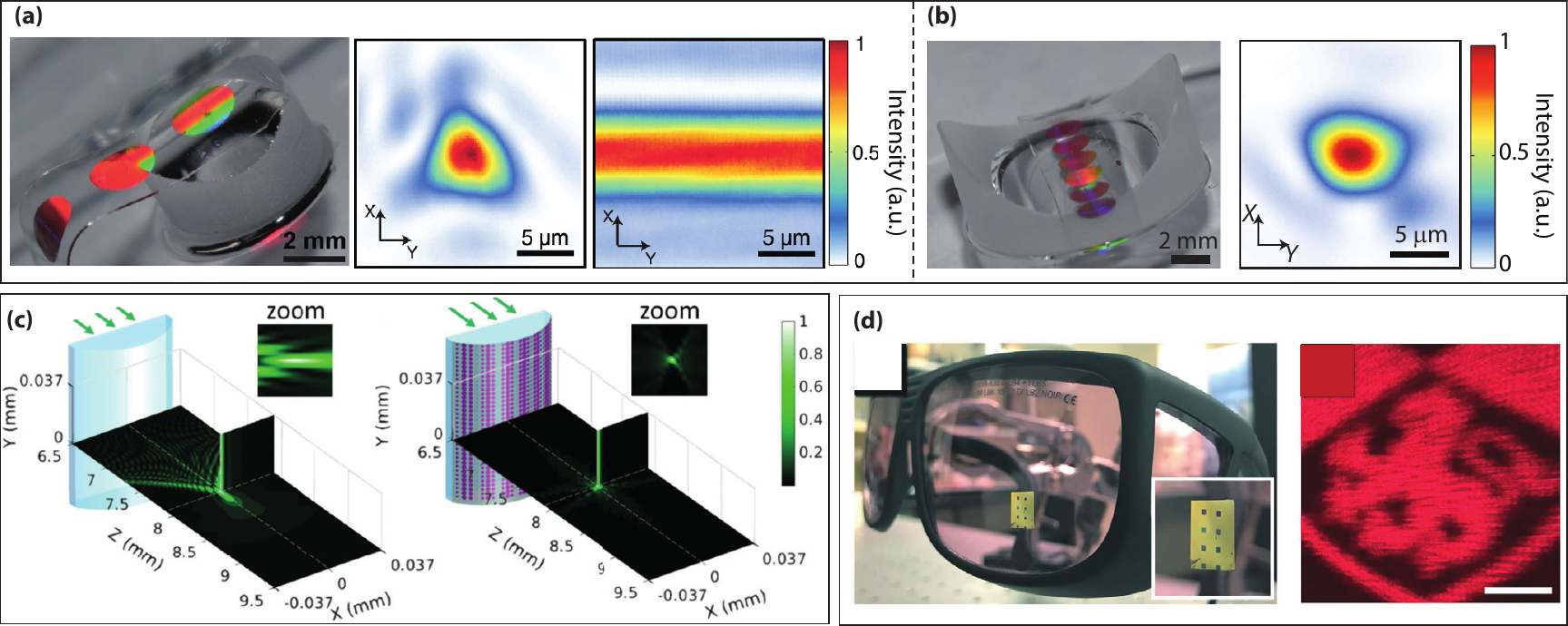}
\caption{\textbf{Conformal metasurfaces}. \textbf{(a)} A conformal dielectric metasurface wrapped over a convex cylinder to make it behave as an aspherical lens. Optical image of the flexible metasurface conformed to the convex glass cylinder is shown on the left. Measured intensities at the plane of focus with (middle) and without (right) the conformal metasurface are also shown~\cite{Kamali2016NatCommun}. \textbf{(b)} Optical image of the fabricated conformal metasurface mounted on a diverging glass cylinder to convert it to an aspherical lens (left). Measured intensity at the designed plane of focus of the metasurface and concave glass cylinder combination under illumination with 915-nm light (right)~\cite{Kamali2016NatCommun}. \textbf{(c)} A conformal metasurface designed to correct spherical aberrations of non-spherical surfaces. Simulated intensity profile of a cylindrical lens is shown on the left. Simulated intensity profile of the metasurface and cylindrical lens combination is shown on the right~\cite{Cheng2016SciRep}. The metasurface is responsible for spherical aberration compensation in the horizontal plane. \textbf{(d)} A conformable metasurface is designed to project a holographic image when mounted on a planar substrate. The degradation of metasurface performance is tolerable with small deformations of the substrate~\cite{Burch2017SciRep}. A conformable metasurface mounted on a slightly curved glass is shown on the left. Measured projected image of the metasurface conformed to the glass is shown on the right.}
\label{fig:8_ConformalMetasurfaces}
\end{figure*}

\begin{figure*}[htp]
\centering
\includegraphics[width=2\columnwidth]{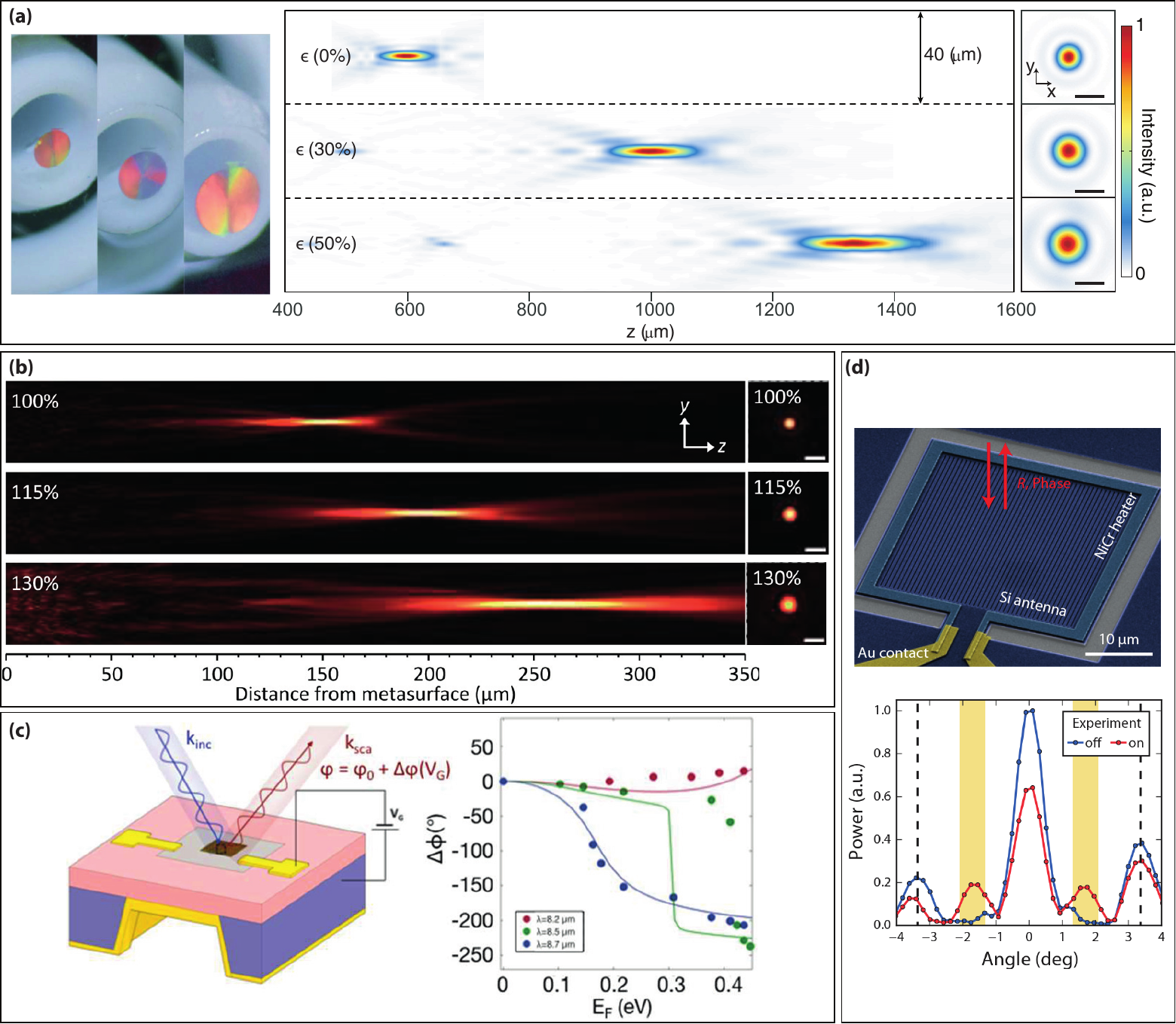}
\caption{\textbf{Tunable metasurfaces}. \textbf{(a)} Highly tunable dielectric metasurfaces based on elastic substrates. Optical images of the tunable elastic metasurface at three different strain values are shown on the left. Measured intensity profiles of the varifocal lens at different strain values in the axial plane and the focal plane are shown on the right~\cite{Kamali2016LaserPhotonRev}. \textbf{(b)} Varifocal lens based on a plasmonic metasurface embedded in an elastic substrate. Measured optical intensity profiles of the elastic metasurface at different strain values in the axial plane (left) and at the corresponding focal planes (right)~\cite{Ee2016NanoLett}. \textbf{(c)} Phase modulation of the reflected light through gate-tunable metasurfaces. Schematic of the tunable device is shown on the left. Reflected light phase modulation for three adjacent wavelengths are shown on the right~\cite{Sherrott2017NanoLett}. \textbf{(d)} Phase-dominant spatial light modulator through thermo-optical effects in a one-sided cavity. False-color scanning electron microscope image of the fabricated single-pixel of the tunable device is shown on the top. Experimental demonstration of beam deflection through a phased array formed from such pixels is shown on the bottom~\cite{Horie2017AcsPhoton}.}
\label{fig:9_TunableMetasurfaces}
\end{figure*}

\begin{figure*}[htp]
\centering
\includegraphics[width=2\columnwidth]{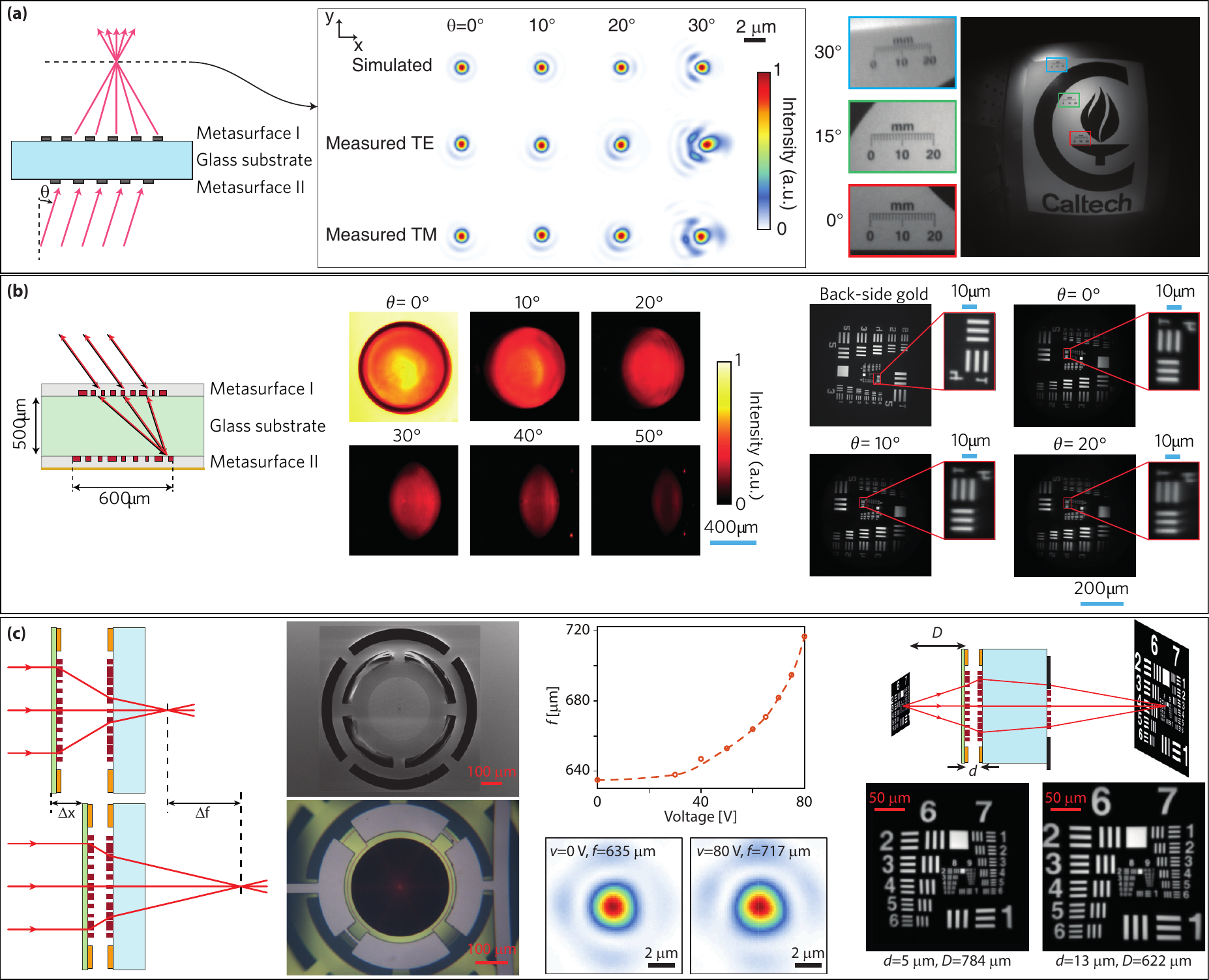}
\caption{\textbf{Metasystems}. \textbf{(a)} A wide-angle metasurface doublet lens. Schematic illustration of corrected focusing by the monolithic metasurface doublet lens, composed of two cascaded metasurfaces is shown on the left. Simulated and measured focal plane intensity profiles of the metasurface doublet for different illumination angles and different polarizations are shown in the middle. The image taken with the metasurface doublet lens, demonstrating the wide-angle operation of the metasystem is shown on the right~\cite{Arbabi2016NatCommun}. \textbf{(b)} Monolithic planar metasurface retroreflector. Schematic drawing of the retroreflector, demonstrating its composition of two cascaded metasurfaces is shown on the left. Measured reflectance of the metasurface retroreflector as a function of the illumination angles is shown in the middle. Measured images of the reflection of an object off the retroreflector as a function of the retroreflector rotation angle is shown on the right~\cite{Arbabi2017NatPhoton}. \textbf{(c)} A micro-electromechanically-tunable metasurface lens doublet~\cite{Arbabi2018NatCommun}. The concept of the MEMS-tunable lens is shown on the left, where changing the separation between metasurfaces tunes the overall focal distance. Scanning electron and microscope images of the device, along with the measured focal distance and intensity distribution in the focal plane are shown in the middle. The concept of an ultra-compact microscope with electrical focusing ability, and simulated imaging results showing the significant movement of the working distance\cite{Arbabi2018NatCommun}.}
\label{fig:Fig10_MetaSystems}
\end{figure*}

\subsection{Wavefront shaping}
One of the important features of metasurfaces is the capability of manipulating the optical wavefront with subwavelength spatial resolution, which enables shaping the wavefront of light with high precision. Therefore, any desired phase profile can be encoded into a metasurface to provide a specific functionality. Lenses \cite{Kato1989ApplOpt,Vo2014IEEEPhotonTechLett,Lalanne2017LaserPhotonRev,
West2014OptExpress,Hasman2003ApplPhysLett,Paniagua2018NanoLett,Markovich2018IEEETAP,
Arbabi2015NatCommun,Klemm2013OptLett}, focusing mirrors \cite{Fattal2010NatPhoton,Arbabi2017Optica,Ni2013LightSciApp,Pors2013NanoLett,
Arbabi2016CLEO_Displess,Yi2017SciRep}, collimators \cite{Arbabi2015OptExp,Yu2010JosaB}, waveplates \cite{Mutlu2012OptExp,Hasman2003ApplPhysLett,Yu2011Science,Yu2012NanoLett,
Huang2012NanoLett,Huang2013NatCommun,Ni2013NatCommun,Lin2014Science}, beam deflectors (gratings) \cite{Lalanne1999JOSAA,Bonod2016AdvanceOptPhoton,Yu2015LaserPhotonRev,
Shalaev2015NanoLett,Zhou2017ACSPhotonics,Sell2017NanoLett,Li2016IEEE,
Huang2014ApplPhysRev,Liu2013ApplPhysLett}, spiral phase plates \cite{Vo2014IEEEPhotonTechLett,Arbabi2015NatNano,Pfeiffer2014PhysRevAppl,
Yang2014NanoLett,Ma2015SciRep,Yi2014OptExpress,Li2015OptExpress}, orbital angular momentum (OAM) generators \cite{Ren2016SciRep,Lin2013NanoLett,Wang2017OptLett,Karimi2014LightSciApp,
Li2013NanoLett,Zhao2017ACSPhoton,Chen2015NanoLett}, and holograms \cite{Kamali2017PRX,Ni2013NatCommun,Wang2016Optica,Zheng2015NatNano,Li2017ACSNano,
Li2017NatCommun,Zhao2016OptLett,Chen2014NanoLett,Wang2016NanoLett,Zhang2016SciRep,Wang2017ACSPhoton} are just some examples of functionalities that are readily available with metasurfaces. Different metasurface platforms have been investigated to create various phase profiles. Although HCAs surpass other proposed platforms in terms of efficiency, robustness, and functionality, all metasurface platforms with meta-atoms that can span the entire 0 to 2$\pi$ phase range can be designed to generate any desired phase profile (with a limited range of possible deflection angles). Figures~\ref{fig:6_WavefrontShaping}a-\ref{fig:6_WavefrontShaping}f demonstrate some examples of wavefront shaping implemented with metasurfaces. Figure~\ref{fig:6_WavefrontShaping}a, shows simulation results of an aspherical metasurface lens designed at a wavelength of 650~nm~\cite{Fattal2011AdvPhoton}. The field distribution of the metasurface lens in the axial plane shows how the wave-front is changed to a spherical profile, causing the light to focus to a diffraction limited spot. The transmission efficiency of the lens is 75$\%$ in simulation. Any other complex wavefronts can be shaped through metasurfaces \cite{Kamali2016NatCommun,Vo2014IEEEPhotonTechLett,Yamada2017OptLett}. Figure~\ref{fig:6_WavefrontShaping}b, shows a scanning electron microscope image of a vortex metasurface lens designed for operation at 850~nm wavelength \cite{Vo2014IEEEPhotonTechLett}. The designed metasurface phase profile is composed of a spherical phase-profile carrying the first order orbital angular momentum (OAM). The measured and simulated efficiency of this metasurface lens are 70$\%$ and 93$\%$, respectively. Metasurfaces can be designed for different wavelengths according to the specific application. Different dielectric and metallic materials have been utilized to demonstrate metasurfaces from visible \cite{Ren2016SciRep,Deng2016OptExp,Yu2015LaserPhotonRev,Karimi2014LightSciApp} to mid-infrared wavelengths \cite{Arbabi2015OptExp,Nordin1999OptExp,Zhang2016OptExp}. With a proper choice of the material system, the design and fabrication of metasrufaces are easily scalable to different wavelengths. For instance, Fig.~\ref{fig:6_WavefrontShaping}c shows a schematic of a mid-infrared metasurface collimating lens~\cite{Arbabi2015OptExp}. The metasurface lens is designed to collimate the output beam radiation of a 4.8-$\mathrm{\mu}$m single-mode quantum cascade laser. The phase distribution at the lens plane is shown in Fig.~\ref{fig:6_WavefrontShaping}c. A 79$\%$ measured transmission efficiency is reported for the metasurface lens with a 0.86 numerical aperture~\cite{Arbabi2015OptExp}. Several dielectric materials have been used for metasurface demonstration at visible, including silicon (in amorphous, poly, or single crystalline form)~\cite{Faraon2016SPIENews,Zhou2017ACSPhotonics,Lin2014Science,Sell2016ACSPhoton}, gallium nitride~\cite{Chen2017NanoLett,Emani2017ApplPhysLett}, silicon nitride~\cite{Ren2016SciRep,Zhan2016ACSPhotonics,Jang2018NatPhoton}, titanium dioxide~\cite{Khorasaninejad2016Science,Lalanne2017LaserPhotonRev,Lalanne1999JOSAA_Multimode}, and silicon dioxide~\cite{Zeitner2012AplPhysA}. Figure~\ref{fig:6_WavefrontShaping}d shows experimental results of a metasurface OAM generator designed at 532-nm wavelength for underwater optical communications. Various blazed grating "fork" phase masks with different OAM orders (+1/-1 and +3/-3) are generated through SiN metasurfaces. Measured intensity profiles and interferograms of +1 and +3 OAM orders are shown in Fig.~\ref{fig:6_WavefrontShaping}d. Phase-only holograms are another examples of phase masks that can be easily demonstrated with metasurfaces. Generally, any target 3D intensity that can be shaped through a digital phase mask, can also be designed with metasurface platforms. Figure~\ref{fig:6_WavefrontShaping}e shows a measured holographic image projected with a metasurface hologram at 1600-nm wavelength with over 90$\%$ transmission efficiency~\cite{Wang2016Optica}. The scanning electron micrograph of a portion of the metasurface is shown in the inset. The thin and planar nature of metasurfaces results in their large angular optical memory effect range~\cite{Feng1988PhysRevLett,Jang2018NatPhoton}. In addition, metasurfaces can also be designed to have large angular scattering ranges (i.e., deflect light to large angles with high efficiency). These characteristics make them suitable for microscopy applications~\cite{Jang2018NatPhoton}. A disorder-engineered metasurface with large angular correlation range ($\sim$30$^\circ$) and a numerical aperture of 0.95 is schematically depicted in Fig.~\ref{fig:6_WavefrontShaping}f. This metasurface along with a spatial light modulator (SLM) is used for complex wavefront engineering to shape diffraction-limited focusing over an extended volume. This combination of SLM and disordered metasurfaces is used for capturing high resolution wide-FOV fluorescence images of biological tissue~\cite{Jang2018NatPhoton}. 

The idea of spatially multiplexing elements to realize multi-functional devices has been widely used in holography and color cameras~\cite{McGrew1983Patent,Bayer1976Patent}. The same concept can also be applied to metasurfaces facilitated by their discrete design nature. Therefore, different phase profiles can be spatially multiplexed to add more functionalities to a device at a cost of efficiency reduction and performance degradation \cite{Huang2015NanoLett,Maguid2016Science,Zhao2016OptLett,Lin2016NanoLett,Zhang2017Annalen}. Figure~\ref{fig:6_WavefrontShaping}g schematically shows two different geometrical multiplexing methods \cite{Maguid2016Science}, named interleaved and harmonic response geometric phase metasurfaces.

Different polarization-dependent and polarization-independent wavefront shaping applications have been demonstrated with metasurfaces. The more interesting capability is the simultaneous control of phase and polarization, which results in realization of novel optical elements \cite{Arbabi2015NatNano,Backlund2016NatPhoton,Mueller2017PhysRevLett,Arbabi2018arXiv}. Figure~\ref{fig:7_PolPhase_application} summarizes a few applications of metasurfaces with polarization-controlled phase, realized through birefringent meta-atoms. For instance, a reflective grating with different deflection angles under \textit{x}- and \textit{y}-polarized light has been realized through birefringent metallic patches at the wavelength of 8.06 $\mathrm{\mu}$m~\cite{Farmahini2013OptLett}. The schematic of one period of the simulated grating, as well as the imposed phases under x and y polarization of light are shown in Fig.~\ref{fig:7_PolPhase_application}a~\cite{Farmahini2013OptLett}. Elliptical $\alpha$-Si nanowires have been utilized to encode different optical functions into different linear polarizations \cite{Schonbrun2011NanoLett}. Figure~\ref{fig:7_PolPhase_application}b demonstrates two-dimensional elliptical nanowire grating designed to diffract \textit{x}-polarized light vertically and \textit{y}-polarized light horizontally~\cite{Schonbrun2011NanoLett}. Diffraction efficiencies of 28.3$\%$ (25.1$\%$) and 27.5$\%$ (26.5$\%$) were reported for the two horizontal (vertical) diffraction orders. With this control of phase and polarization, different wavefronts can be embedded in a single metasurface and separately retrieved under different polarizations. A polarization switchable hologram that generates two different images for \textit{x}- and \textit{y}-polarized light is shown in Fig.~\ref{fig:7_PolPhase_application}c~\cite{Arbabi2015NatNano}. Measured efficiencies of 84$\%$ and 91$\%$ were reported for this device for \textit{x}- and \textit{y}-polarized incident light, respectively. Recently, the dielectric metasurfaces with the ability to control phase and polarization were utilized to demonstrate full-Stokes polarization cameras~\cite{Arbabi2018arXiv}. This category of metasurfaces has also been employed in microscopy applications~\cite{Backlund2016NatPhoton}. A metasurface that converts azimuthally polarized light into \textit{y}-polarized light has been used to increase the accuracy of molecular localization in super-resolution imaging techniques. Experimental results, showing the higher localization accuracy with the metasurface mask is shown in Fig.~\ref{fig:7_PolPhase_application}d.

\subsection{Conformal metasurfaces}
The ultrathin thickness and 2D nature of metasurfaces make them suitable for transferring to flexible substrates that can conform onto the surface of nonplanar objects to change their optical properties. Flexible metasurfaces with this capability of breaking the correlation between the geometry of an object and its optical functionality are called conformal metasurfaces~\cite{Kamali2016NatCommun,Cheng2016SciRep,Germain2013APL,Jiang2017NatCommun,Burch2018ACSPhoton}. Conformal metasurfaces have applications where a specific optical functionality must be provided while the shape is dictated by other considerations. Phase cloaking is just one example of functionalities enabled by conformal metasurfaces~\cite{Chen2013PhysRevLett,Orazbayev2017AdvOptMater}.  Several efforts have been made to transfer metasurfaces (often plasmonic devices) to flexible and elastic substrates, mostly with the aim of frequency response tuning through substrate deformation~\cite{Reyce2010NanoLett,Xu2011NanoLett,DiFalco2011APL,Walia2015AplPhysRev,Gutruf2016ACSNano,Zhu2015Optica}. Most of the demonstrated platforms are at longer wavelengths than in the optical regime, because transferring metallic and large structures to flexible substrates is more feasible~\cite{Srivastava2017APL,Cong2014LPR,Germain2013APL,Zhang2016SciRep}. The idea of conformal metasurfaces was first proposed and demonstrated experimentally through HCAs for the optical regime in~\cite{Kamali2016NatCommun}. Two different examples of dielectric metasurfaces wrapped over cylindrical surfaces to convert them to aspherical lenses are illustrated in Figs.~\ref{fig:8_ConformalMetasurfaces}a and \ref{fig:8_ConformalMetasurfaces}b~\cite{Kamali2016NatCommun}. Figure~\ref{fig:8_ConformalMetasurfaces}a shows a flexible metasurface conformed to a convex glass cylinder to make it behave like a converging aspherical lens. Intensities at the focal plane with and without the metasurface demonstrate a dramatic change in the optical behavior of the nonplanar object. Another demonstration that showcases the capability of this platform is shown in Fig.~\ref{fig:8_ConformalMetasurfaces}b, where a metasurface is wrapped over a diverging glass cylinder to make it behave like a converging aspherical lens. A very robust fabrication process with a near-unity yield (larger than 99.5 $\%$ reported) has been developed for transferring large areas of dielectric metasurfaces (centimeter scales) to flexible substrates~\cite{Kamali2016NatCommun}. It is worth noting that the alignment of a conformal metasurface to a non-planar substrate needs to be very accurate for achieving the optimal phase compensation through the metasurface. Different groups have investigated the realization of conformal metasurfaces~\cite{Cheng2016SciRep,Burch2017SciRep,Burch2018ACSPhoton}. Figure~\ref{fig:8_ConformalMetasurfaces}c shows a simulation for a proof of concept application of conformal metasurfaces for spherical aberration correction~\cite{Cheng2016SciRep}. Figure~\ref{fig:8_ConformalMetasurfaces}d shows a metasurface conformed to a glass surface with a large radius of curvature that projects a holographic image~\cite{Burch2017SciRep}. The metasurface itself is designed for a flat surface and provides its best functionality on a flat surface. The degradation resulting from its conformation on a curved surface is tolerable if the radius of curvature of the object is much larger than the curvature of the wavefront generated by the metasurface.

\subsection{Tunable metasurface optical devices}
The properties of all of the metasurface devices discussed above are generally fixed. It is highly desirable to tune, switch, or reconfigure the functionality of these optical elements. Different approaches have been proposed for building tunable metasurfaces such as mechanical deformations and reconfigurations~\cite{Kamali2016LaserPhotonRev,Ee2016NanoLett,She2018SciAdv,Zhan2017SciRep,Tseng2017NanoLett}, electrical and magnetic tuning~\cite{Ou2013NatNanotech,Sherrott2017NanoLett,Huang2016NanoLett,Iyer2016AdvOptMat,Colburn2017SciRep,Fallahi2012PhysRevB,
Yao2014NanoLett,Kim2017OptLett,Komar2017ApplPhysLett,Sautter2015NanoLett,Bar-David2017NanoLett}, thermal tuning \cite{Horie2017AcsPhoton,Ou2011NanoLett,Colburn2017SciRep}, and material reconfiguration through phase-change materials \cite{Hosseini2017Materials,Michel2013NanoLett,Wutting2017NatPhoton}.

Mechanically tunable metasurfaces based on elastic substrates are among the promising tuning platforms that have enabled varifocal lenses~\cite{Kamali2016LaserPhotonRev,Ee2016NanoLett}, color tuning~\cite{Zhu2015Optica}, and frequency response tuning~\cite{Gutruf2016ACSNano}. Figure~\ref{fig:9_TunableMetasurfaces}a shows a varifocal aspherical metasurface lens tuned through radially stretching a thin substrate ($\sim$100-$\mathrm{\mu}$m thick) that embeds the metasurface layer. A large tuning range (over 952 diopters change in the optical power), while maintaining high efficiency (above 50$\%$ focusing efficiency) and polarization independent performance is reported in~\cite{Kamali2016LaserPhotonRev}. Another demonstration of varifocal lenses through elastic plasmonic metasurfaces is shown in Fig.~\ref{fig:9_TunableMetasurfaces}b~\cite{Ee2016NanoLett}. The metasurface is designed using the geometric phase, therefore it works for one circular polarization. Moreover, electrically tunable elastomers can be utilized to actuate the same varifocal lenses electrically~\cite{She2018SciAdv}. Mechanical movements of multiple rigid metasurfaces have been also investigated to provide focal distance tuning through demonstration of Alvarez lenses~\cite{Zhan2017SciRep}.

Electrically driven carrier accumulation is another proposed method of phase modulation in metasurfaces~\cite{Sherrott2017NanoLett,Huang2016NanoLett,Park2017NanoLett}. Figure~\ref{fig:9_TunableMetasurfaces}c shows a gate-tunable graphene-gold resonator geometry, with more than 230$^\circ$ reflected phase modulation range. A calculated efficiency of about 1$\%$ is reported for this device. Electrically driven liquid crystals have also been used for frequency response modulation of metasurfaces~\cite{Kim2017OptLett,Sautter2015NanoLett}. Electrostatic forces can also be utilized for tuning the frequency response of metasurfaces~\cite{Burokur2010AppPhysLett}.

Thermal tuning is another method of changing the response of metasurfaces. Thermo-optical modulation of metasurfaces have been proposed for spatial light modulation \cite{Horie2017AcsPhoton,Donner2015ACSPhotonics}. Figure~\ref{fig:9_TunableMetasurfaces}d shows a high-speed silicon-based device for phase-dominant spatial light modulation through the thermo-optic effect in $\alpha$-Si.
The building block of the device is composed of an asymmetric Fabry-P\'{e}rot resonator formed by a silicon subwavelength grating reflector and a distributed Bragg reflector. The pixels exhibit nearly 2$\pi$ phase-dominant modulation with a speed of tens of kHz at telecom wavelengths~\cite{Horie2017AcsPhoton} that can be utilized for beam steering. Frequency response tuning of metasurfaces has been also demonstrated through thermo-mechanical actuation~\cite{Ou2011NanoLett}. Another tuning method is based on micro-electromechanical systems (MEMS)~\cite{Yoo2014OptExp,Arbabi2018NatCommun}. For instance, Yoo et al. demonstrated a MEMS-based phased array using moving high-contrast grating mirrors~\cite{Yoo2014OptExp}. 

Other platforms have also been proposed for implementing reconfigurable and switchable metasurfaces. For instance, reconfigurable metasurfaces can be achieved through phase-change materials~\cite{Hosseini2017Materials,Michel2013NanoLett,Wutting2017NatPhoton,Chu2016LaserPhotonRev}. The optical properties of phase change materials vary through transformation from the amorphous to the crystalline state. Usually short optical or electrical pulses are utilized for actuating these materials to switch between the two states to provide reconfigurable metasurfaces~\cite{Wutting2017NatPhoton}. It is noteworthy that metasurfaces with angular response control~\cite{Kamali2017PRX} and polarization-phase control~\cite{Arbabi2015NatNano} that were discussed above, can also be considered as switchable metasurfaces where the switching of the response is achieved by changing the incident and angle or polarization of the incident light, respectively. These platforms enable large changes in the response of the metasurface as a function of the incident angle or polarization.

The required power or voltage (in the case of electrical tuning) might be another important factor in tunable and reconfigurable metasurfaces. Comparison of the required tuning power for various systems is complicated as different works report these values for different tuning modalities (e.g., total system function tuning versus pixel tuning, or high speed modulation versus static reconfigurability, etc.). In addition, in practice the required power for the driving electronics should also be taken into account. Nevertheless, various experiments have been reported demonstrating pixel tuning with millivolt-~\cite{Thyagarajan2017AdvMat} and milliwatt-scale~\cite{Horie2017AcsPhoton} actuation.

\subsection{Metasystems}
In this section we review recent advances of metasurfaces as building blocks of  metasystems. By metasystems, we refer to specific arrangements of multiple metasurfaces designed to provide a functionality that is not possible to achieve with a single metasurface layer. Metasurfaces with the capability of providing sophisticated planar wavefronts are very suitable to be vertically integrated through monolithic processes, thus creating high-performance and low-cost miniature optical systems. These optical systems can be directly integrated with image sensors and other optoelectronic elements without the need for post-fabrication alignments, thus enabling low-power and low-weight miniature optical systems. Figure~\ref{fig:Fig10_MetaSystems} summarizes some of the recent demonstrations of the meta-systems.

A miniature planar camera, shown in Fig.~\ref{fig:Fig10_MetaSystems}, was implemented through the integration of a metasurface doublet corrected for monochromatic aberrations. A doublet lens was formed by cascading two metasurfaces with a glass spacer layer in between~\cite{Arbabi2016NatCommun}. The phase profiles of the two metasurfaces were optimized to provide near-diffraction-limited focusing for a wide range of incident angles (a field of view larger than 60$^\circ\times$60$^\circ$). The doublet acts as a fisheye photographic objective operating at a wavelength of 850 nm with a small f-number of 0.9, and measured focusing efficiency of 70$\%$ under normal illumination. It is notable that a singlet metasurface lens with similar focal distance and f-number exhibits significant aberrations even at incident angles of a few degrees, while the demonstrated metasystem (doublet lens) provides a nearly diffraction-limited focal spot for incident angles up to more than 25$^\circ$~\cite{Arbabi2016NatCommun}. More recently, the same concept and design have been implemented in the visible region (at 532~nm) using geometric phase TiO$_2$ HCAs. In the latter device, a field of view of 50$^\circ$ and a maximum focusing efficiency of 50$\%$ operating under one circular polarization were obtained~\cite{Groever2017NanoLett}.

Later, a planar monolithic metasurface retroreflector was demonstrated using two vertically stacked metasurfaces~\cite{Arbabi2014CLEO_Retro,Arbabi2017NatPhoton}. The first metasurface performs a spatial Fourier transform and maps light with different illumination angles to different spots on the second metasurface, while the second metasurface adds a spatially varying momentum to the incoming light. The demonstrated metasystem (retroreflector) reflects light along its incident direction with a large half-power field of view of 60$^\circ$. A measured efficiency of 78$\%$ was reported under normal illumination~\cite{Arbabi2017NatPhoton}. We should note here that it is fundamentally impossible to make a retroreflector operating over a wide continuous range of angles with a single layer of a local metasurface~\cite{Kamali2017PRX,Kamali2017SPIEPW,Jang2018NatPhoton}.

Very recently, integration of metasurfaces with the micro-electromechanical systems (MEMS) has been utilized to demonstrate tunable lenses [Fig.~\ref{fig:Fig10_MetaSystems}c]. The concept of this tunable lens is shown in Fig.~\ref{fig:Fig10_MetaSystems}c (left), where changing the separation of the metasurface lenses by a small distance ($\Delta$x$\sim$1~$\mu$m) can tune the overall focal length by a significantly larger amount ($\Delta$f$\sim$30 $\mu$m). Scanning electron and microscope images of the fabricated device, along with the measured focal distances and in-focus intensity distributions are plotted in Fig.~\ref{fig:Fig10_MetaSystems}c, middle. A similar tunable metasurface doublet is combined with a third metasurface that is patterned on the other side of the glass substrate to demonstrate an ultra-compact microscope ($\sim$1~mm$^3$) [Fig.~\ref{fig:Fig10_MetaSystems}c, right]. The microscope is designed to have a large corrected field of view ($\sim$500~$\mu$m, or about 40 degrees), while its working distance can be electrically tuned over 150~$\mu$m.

\section{Outlook: potentials and challenges}
We briefly discussed different proposed metasurface platforms and their functionalities. Among various platforms investigated so far, HCAs outperform the other ones in wavefront manipulation as they provide high efficiencies and novel functionalities such as control over the polarization, spectral, and angular degrees of freedom that are not available using other platforms. The development and study of optical metasurfaces have been a rapidly growing field of research in the past few years, because of their capabilities to mimic the functionality of conventional diffractive optical elements with higher efficiencies and resolutions, and more importantly for their advantages in providing new functionalities not achievable with conventional diffractive optics. Their subwavelength thickness, planar form factor, compatibility with conventional micro/nano-fabrication techniques, potentially low-cost batch fabrication, ability to replace a system of multiple bulky conventional elements with a miniature element, new capabilities to control different degrees of freedom of light, and prospects for a paradigm change in how optical systems are designed, make them very promising for the realization of the next generation of compact high-performance optical systems.

Despite all the advancements made in the past few years, several challenges still remain unresolved both from fundamental and practical points of view. An important theoretical issue is the number of available degrees of freedom that exist in a single surface or a specific volume. This would determine the number of functionalities that can be encoded in such a device with negligible performance degradation. The importance of this issue becomes more clear as one considers the great interest in realizing multi-functional metasurfaces. Despite several such devices including multi-wavelength metasurfaces~\cite{Khorasaninejad2015NanoLett,Arbabi2016Optica,Arbabi2016OptExp}, multi-angle metasurfaces~\cite{Kamali2017PRX,Cheng2017SciRep}, and metasurfaces with independent polarization and phase control~\cite{Arbabi2015NatNano,Mueller2017PhysRevLett}, the number of available degrees of freedom in such a device, and how exactly they can be utilized is still mostly unknown. Although optimization techniques have been used to improve the performance of multi-functional devices~\cite{Sell2017NanoLett,Sell2017AdvOptMat,Lin2017arXiv}, they still do not determine the possible number of functionalities. Another area which requires significant advancements is the modeling and design of non-periodic metasurfaces. Currently, almost all of the design methods are based on results of simulation of periodic lattices of the meta-atoms. Although this approach works well for slowly varying metasurfaces with small deflection angles, its underlying assumptions (namely locality, angle independence, and weak coupling of meta-atoms) cease to be valid for devices with large deflection angles. Therefore, more precise design methods that take all of these into account, and at the same time can be applied to large non-periodic structures are of great interest. In addition to enabling high-efficiency high-NA devices, such methods could also allow for the design and analysis of novel metasurfaces that are not bound by the assumption of locality. Finally, despite several attempts at realizing achromatic and dispersion-engineered metasurfaces, the operation bandwidths, sizes, and numerical apertures of devices that are possible with the existing platforms are very limited. The dependence of all of these limitations on the possible controllable quality factors that the platform provides makes the problem even more challenging. As a result, there is still a long way to the realization of achromatic and dispersion-engineered metasurfaces with practical sizes (i.e., aperture sizes of a few millimeters) and moderate to high numerical apertures.

In addition to fundamental challenges, there are also several unresolved practical issues hindering the realization of high-volume low-cost metasurface devices for real-life applications. One issue worth addressing is the absence of a low-loss high-index material for visible light. Although there have been several realization of dielectric metasurfaces in the visible~\cite{Ren2016SciRep,Deng2016OptExp,Emani2017ApplPhysLett,Khorasaninejad2016NanoLett}, their efficiency is still not as high as infrared metasurfaces where materials with low loss and high refractive index like silicon can be used.  This is especially true for the cases of polarization independent metasurfaces, and devices with independent control of phase and polarization. In addition, for several applications, it is essential that the metasurface is capped by a low-index material (for instance for mechanical robustness, fabrication requirements, or realization of flexible and conformal metasurfaces). In such scenarios, the refractive index of currently available low-loss materials in the visible is not high enough to provide low-coupling between nanoposts and full phase coverage.

To have a significant industrial impact, the manufacturing processes of metasurfaces should be compatible with the existing low-cost large-scale foundry technology. Although this might already be possible for devices working in near and mid-IR (above 1.5~$\mu$m wavelength), it is challenging for devices that work below 1~$\mu$m, which are fabricated almost exclusively with electron beam litography. In principle, large-scale fabrication techniques like deep UV lithography, roll-to-roll nanoimprint, and soft lithography could address this challenge; however, there still exist practical barriers that should be overcome before this becomes a reality.

Another category of highly desirable devices is the tunable metasurfaces. Despite several demonstrations of wavefront tuning using metasurfaces, none of them can still compete with the commercially available liquid crystal based spatial light modulators. High-efficiency, ultrafast, high-resolution wavefront tuning is of great need, and there is a lot of room for optimizing high-performance metadevices for beam steering applications, spatial light modulators, and dynamic holographic displays.

With the future advancements of metasurfaces in mind, we envision them at least as a complementary platform, if not a paradigm changing one, in optical element and system design for various applications.

\bibliographystyle{naturemag_noURL}

\bibliography{MetasurfaceLibrary}

\end{document}